\documentclass[%
 reprint,
superscriptaddress,
nofootinbib,
 amsmath,amssymb,amsart,
prd,
tightenlines
]{revtex4-2}

\usepackage{graphicx}% Include figure files
\usepackage{dcolumn}% Align table columns on decimal point
\usepackage{bm}% bold math
\usepackage{color}
\definecolor{linkcolor}{rgb}{0.6,0,0}
\definecolor{citecolor}{rgb}{0,0,0.75}
\definecolor{urlcolor}{rgb}{0.12,0.46,0.7}
\usepackage[breaklinks, colorlinks, urlcolor=urlcolor, linkcolor=linkcolor,citecolor=citecolor,pdfencoding=auto]{hyperref}
\usepackage{amsmath}
\usepackage{ulem}
\usepackage{xspace}
\usepackage{booktabs}
\usepackage{xcolor,colortbl}

\newcommand{\planck}{\textit{Planck}\xspace}

\newcommand{\xqml}{\texttt{xQML}\xspace}
\newcommand{\nver}{\hat{\mathbf{n}}}
\renewcommand{\textdegree}{$^\circ$ }
\newcommand{\clmumu}{C_\ell^{\mu\mu}\xspace}
\newcommand{\clmut}{C_\ell^{\mu T}\xspace}
\newcommand{\clmue}{C_\ell^{\mu E}\xspace}
\newcommand{\clmub}{C_\ell^{\mu B}\xspace}
\newcommand{\fsky}{f_{\mathrm{sky}}\xspace}
\newcommand{\fnl}{f_{\rm NL}\xspace} %f^{\rm loc}_{\rm NL}
\newcommand{\nnl}{n_{\rm NL}\xspace}

\newcommand{\ks}{KS15\xspace}
\newcommand{\fixsenspec}{F96}

\newcommand{\fnlT}{\ensuremath{5.7 \times 10^6}}
\newcommand{\fnlE}{\ensuremath{5.8 \times 10^6}}
\newcommand{\fnlTE}{\ensuremath{3.6 \times 10^6}}

\begin{document}

\preprint{}

\title{CMB spectral distortions revisited: a new take on $\mu$ distortions and primordial non-Gaussianities from FIRAS data}

\author{F. Bianchini}%
\email{fbianc@stanford.edu}
\affiliation{Kavli Institute for Particle Astrophysics and Cosmology, Stanford University, 452 Lomita Mall, Stanford, CA, 94305, USA}
\affiliation{SLAC National Accelerator Laboratory, 2575 Sand Hill Road, Menlo Park, CA, 94025, USA}
\affiliation{Department of Physics, Stanford University, 382 Via Pueblo Mall, Stanford, CA, 94305, USA}
\author{G. Fabbian}%
\email{gfabbian-visitor@flatironinstitute.org}
\affiliation{Center for Computational Astrophysics, Flatiron Institute, New York, NY 10010, USA}
\affiliation{School of Physics and Astronomy, Cardiff University, The Parade, Cardiff, CF24 3AA, UK}

\date{\today}

\begin{abstract}
Deviations from the blackbody spectral energy distribution of the Cosmic Microwave Background (CMB) are a precise probe of physical processes active both in the early universe (such as those connected to particle decays and inflation) and at later times  (e.g. reionization and astrophysical emissions). Limited progress has been made in the characterization of these spectral distortions after the pioneering measurements of the FIRAS instrument on the COBE satellite in the early 1990s, which mainly targeted the measurement of their average amplitude across the sky. Since at present no follow-up mission is scheduled to update the FIRAS measurement, in this work we re-analyze the FIRAS data and produce a map of $\mu$-type spectral distortion across the sky. We provide an updated constraint on the $\mu$ distortion monopole $|\langle\mu\rangle|<47\times 10^{-6}$ at 95\% confidence level that sharpens the previous FIRAS estimate by a factor of $\sim 2$. We also constrain primordial non-Gaussianities of curvature perturbations on scales $10\lesssim k\lesssim 5\times 10^4$ through the cross-correlation of $\mu$ distortion anisotropies with CMB temperature and, for the first time, the full set of polarization anisotropies from the \planck satellite. We obtain upper limits on $\fnl\lesssim \fnlTE$ and on its running $\nnl\lesssim 1.4$ that are limited by the FIRAS sensitivity but robust against galactic and extragalactic foreground contaminations. We revisit previous similar analyses based on data of the \planck  satellite and show that, despite their significantly lower noise, they yield similar or worse results to ours once all the instrumental and astrophysical uncertainties are properly accounted for. Our work is the first to self-consistently analyze data from a spectrometer and demonstrate the power of such instrument to carry out this kind of science case with reduced systematic uncertainties.

\end{abstract}

\maketitle

% \tableofcontents

\section{\label{sec:intro}Introduction} 
Observations of the cosmic microwave background (CMB) arguably represent the cornerstone of modern cosmology.
Over the last three decades, accurate measurements of the CMB temperature and polarization anisotropies have provided us with a snapshot of the universe at the time of recombination and have yielded stringent constraints on the constituents, dynamics, and geometry of the universe \citep{planck18_params,aiola20,polarbear20,dutcher21}.
This picture is complemented by measurements of the intensity spectrum of the CMB which directly probe the thermal history of the universe, providing access to additional cosmological information not encoded in the spatial anisotropies \citep{mather93,zannoni08,fixsen11}. 
In particular, departures from a pure blackbody spectral energy distribution, the so-called "spectral distortions" (SD), open a special window on the physical processes active before recombination as well as at more recent times \citep[e.g.,][]{chluba19}.

SD naturally arise when thermalization is inefficient in keeping matter and radiation in thermodynamical equilibrium. 
Examples of mechanisms that can drive the photo-baryonic fluid out of equilibrium are dissipation of primordial acoustic waves and energy injections in the form of photons or other electromagnetically-interacting particles (see e.g. \citep{Lucca2019,Fu2020} and references therein for a recent review). 
Based on their spectral shape, we can broadly break down CMB SD into two classes: the $\mu$- and $y$-type distortions.
At redshifts greater than $z_{\mu}\gtrsim 2\times 10^{6}$, the creation and redistribution of photons by Compton scattering, double Compton scattering, and bremsshtrahlung restore thermal equilibrium and erase any SD.
Between $ 5\times 10^4  \lesssim z  \lesssim 2\times 10^6$, when double Compton scattering and bremsshtrahlung become inefficient, energy injections result in a Bose-Einstein distribution with a non-zero effective chemical potential $\mu$, giving rise to $\mu$-type distortions \citep[e.g.,][]{sz70,burigana91}. 
Around $z_{\mu y} \lesssim 5\times10^4$ Compton scattering becomes inefficient too and photons fall out of kinetic equilibrium with electrons, sourcing the $y$-type distortions.
In reality, the transition from the $\mu$- to the $y$-era is not abrupt and as a result, residual distortions at intermediate times which cannot be fully described by the sum of $\mu$- and $y$-types (the so-called $r$-type distortions) form. Their magnitude within $\Lambda$CDM is however expected to be smaller than the sensitivity of proposed future missions like PIXIE \citep[e.g.,][]{chluba14}.
The SD sensitivity to different cosmic epochs not only allows us to probe standard and exotic physics, but also the power spectrum of primordial fluctuations over a broad range of scales, $1 \lesssim k \lesssim 10^4$ Mpc$^{-1}$.
While $y$-type distortions are generated both in the early universe and during the reionization and structure formation epochs \citep[e.g.,][]{hu94,refregier00,mroczkowski19}, $\mu$-type distortions are only produced in the pre-recombination era, making them a clean and powerful probe of the early universe physics. 
For this reason, we shall focus on the latter type of distortions in this paper.

Observational bounds on the monopole component of these distortions have been set by COBE/FIRAS and are at the level of $|\langle\mu\rangle|<90\times10^{-6}$ and $|\langle y\rangle|<15 \times 10^{-6}$ (95\% CL) \citep{mather93,fixsen96}. 
At lower frequencies than those covered by FIRAS, ARCADE \citep{fixsen11} and TRIS \citep{zannoni08} have more recently carried out absolute measurements of the CMB spectrum and derived similar constraints. 
Several follow-up experiments like PIXIE, PRISM, BISOU, COSMO  \citep{prism,pixie2016,bisou,cosmo} have been proposed to improve absolute measurements of the CMB spectrum. 

Apart from the Sunyaev-Zel'dovich (SZ) distortion sourced by galaxy clusters on arcminute scales \citep[e.g.,][]{zeldovich69,carlstrom02,mroczkowski19}, CMB SD are predicted to be isotropic signals in the simplest cosmological scenarios.
There is, however, an intriguing possibility. 
If the spectrum of primordial perturbations is non-Gaussian, then a spatial modulation of the SD will be induced, leading to a potentially observable anisotropic pattern of the distortions, as first discussed in \citet{pajer12}.
Specifically, the local-type non-Gaussianity which peaks in the so-called squeezed configuration\footnote{Defined as the bispectrum configuration where two wavenumbers are much larger than the third one, $k_1\ll k_2 \simeq k_3$.} correlates the small-scale primordial power spectrum, traced by the SD, with the long-wavelength modes probed, e.g., by primary CMB, inducing a non-zero cross-correlation between $\mu$ and $T$, $E$-, and $B$-modes  \citep[e.g.,][]{pajer12,ganc12,emami15,ota16,ravenni17,orlando21}.
As such, a measurement of the spatial correlation between the $\mu$-type distortion and primary CMB anisotropies can constrain the amplitude of primordial non-Gaussianity (encoded by the dimensionless parameter $\fnl$) at very small scales with wavenumbers of about $k\simeq 10^2-10^4$ Mpc$^{-1}$.
This information would greatly complement the view on primordial local non-Gaussianities from \textit{Planck}'s measurement of $T$ and $E$ anisotropies bispectrum at much larger scales ($\sigma(\fnl)\simeq 5$ at $k \lesssim 0.15$ Mpc$^{-1}$ \citep{planck18_ng}), providing key insights on the scale-dependence of the primordial non-Gaussian signal and hence on the inflationary mechanism.

In this paper, we use archival data from the COBE/FIRAS experiment to reconstruct a sky map of the $\mu$-type distortion fluctuations. 
We then correlate this map with the primary CMB temperature and, for the first time, the full set of polarization anisotropies observed by the \planck satellite. 
The extracted SD-CMB cross-power spectra are then translated into constraints on the amplitude of primodial non-Gaussianity of the local-type, $\fnl^{\rm loc}$ (we will drop the superscript hereafter for simplicity).

We point out that reconstructing the $\mu$-type distortion anisotropies does not necessarily require an absolute measurement of the sky.
The work of \citet{khatri15} was, to our knowledge, the first one that attempted a reconstruction of the fluctuating part of the $\mu$ distortions using a component separation method applied to imager data, namely those from the high-frequency instrument on board of the \planck satellite. 
However, this approach can be more affected by contaminations from residual primary CMB and other astrophysical foregrounds \citep[e.g.,][]{remazeilles18,remazeilles21}.
In addition, we stress that knowledge of the $\mu$ monopole is needed to break the degeneracy between the average level of $\mu$ distortions, $\langle\mu\rangle$, and $\fnl$ (see \citep{ganc12,khatri15,remazeilles18,mukherjee19,remazeilles21} for discussions on measuring $\mu$ fluctuations with a relatively calibrated experiment). This possibility is only allowed by instruments such as spectrometers that are sensitive to the absolute sky temperature.

The paper is organized as follows. In Sec.~\ref{sec:data} we briefly introduce the main ingredients of this analysis, the COBE/FIRAS and \planck datasets, while the analysis methodology, from the sky modelling to the cosmological inference, is reviewed in Sec.~\ref{sec:methods}. Maps of the anisotropic $\mu$-type distortion, its cross-correlation with primary CMB anisotropies, and systematic checks are presented in Sec.~\ref{sec:results}. We discuss constraints on primordial non-Gaussianities in Sec.~\ref{sec:fNL_constaints} and compare to previous measurements in Sec.~\ref{sec:khatri}. Finally, we draw our conclusions in Sec.~\ref{sec:conclusions}.

\section{\label{sec:data}datasets} 
\subsection{\label{sec:firas}COBE/FIRAS} 
The Far Infrared Absolute Spectrophotometer (FIRAS) instrument was a cryogenically cooled Martin-Puplett interferometer on-board of the Cosmic Background Explorer (COBE) satellite.
Designed to cover the frequency range between 30  GHz to 2910  GHz in two spectral bands, FIRAS was able to provide accurate measurements of the CMB spectral energy distribution, thermal emission from interstellar dust, and various infrared cooling lines of the interstellar gas  \citep{mather93}.
To accomplish this, FIRAS used a differential optical system with two inputs, one collecting radiation from the sky and one from an internal reference calibrator, and two outputs feeding radiation to composite bolometer detectors. 
The required absolute calibration was achieved through an external blackbody calibrator.
The FIRAS horn antenna accepted incoming sky radiation from a 7\textdegree circle, resulting in an approximately top-hat beam function.

In this work we use the calibrated FIRAS skymaps of spectra that have had post-calibration offsets removed, a process known as ``destriping".
Specifically, we focus on the final delivery of the FIRAS low frequency low spectral resolution destriped sky spectra, which cover the spectral range from 60 up to 630  GHz in frequency bins with a width of about $\Delta\nu\approx 13$  GHz. The twofold reasons for doing so are that the CMB emission becomes almost entirely negligible at higher frequencies while the noise levels significantly increase.
The relevant data products are provided as maps in \textsc{HEALPix} pixelization\footnote{\url{http://healpix.sourceforge.net}} at a resolution of $N_{\rm nside}=16$, corresponding to an approximate pixel size of $\delta\theta \sim 3.5^\circ$\footnote{\url{https://lambda.gsfc.nasa.gov/product/cobe/firas_tpp_all_get.cfm}}, for each observed frequency.
For a detailed discussion on the instrument, the calibration process, and the released data products, we refer the reader to the FIRAS Explanatory Supplement.\footnote{\url{https://lambda.gsfc.nasa.gov/product/cobe/firas_exsupv4.cfm}}

\subsection{\label{sec:planckdata}\planck CMB maps} 
The \planck satellite was the third generation of space-based missions, after COBE and WMAP, to study CMB physics and was dedicated to image the temperature and polarization CMB anisotropies  at high angular resolution.
Launched in 2009, about 20 years after COBE, \planck carried out a full-sky survey of the microwave sky in nine bands across the 30 to 857  GHz range, down to $\sim 5^\prime$ angular resolution \citep{planck18_overview}. The broad spectral coverage of Planck enabled an accurate characterization and separation of the diffuse foregrounds. 
Unlike FIRAS, \planck is not sensitive to the absolute brightness temperature of the sky. Its calibration thus assumes the knowledge of the CMB temperature, as measured by FIRAS, and relies on the modulation of the CMB dipole anisotropy induced by the motion of the spacecraft with respect to the CMB reference frame \cite{planck18_lfi,planck18_hfi}. 
The \planck maps from the 2018 release are calibrated with a precision approaching $\sim 10^{-4}$ at frequencies below 500 GHz and more recent analyses extended this calibration procedure at the highest frequency channels \cite{planck20_npipe}. This accurate measurement in multiple frequency channels allowed to separate the CMB from other astrophysical emissions using different approaches.

In this analysis, we use four foreground-cleaned CMB temperature and polarization anisotropies maps publicly released by the \planck collaboration and derived with different component separation algorithm in both real and harmonic domain:  Commander, Needlet Internal Linear Combination (NILC), Spectral Estimation Via Expectation Maximization (SEVEM), and Spectral Matching Independent Component Analysis (SMICA) \cite{planck18_compsep}.
These maps are also provided in the \textsc{HEALPix} format at  $N_{\rm nside}=2048$. We first convolve the \planck component-separated maps with the FIRAS scanning beam (which includes the instantaneous optical response and its variation due to the satellite motion as described in Sec.~\ref{sec:ps_ext}) and then downgrade them to a resolution of $N_{\rm nside}=16$ to produce maps at the native FIRAS resolution.

\subsection{Sky masks}
\begin{figure}
\includegraphics[width=\columnwidth]{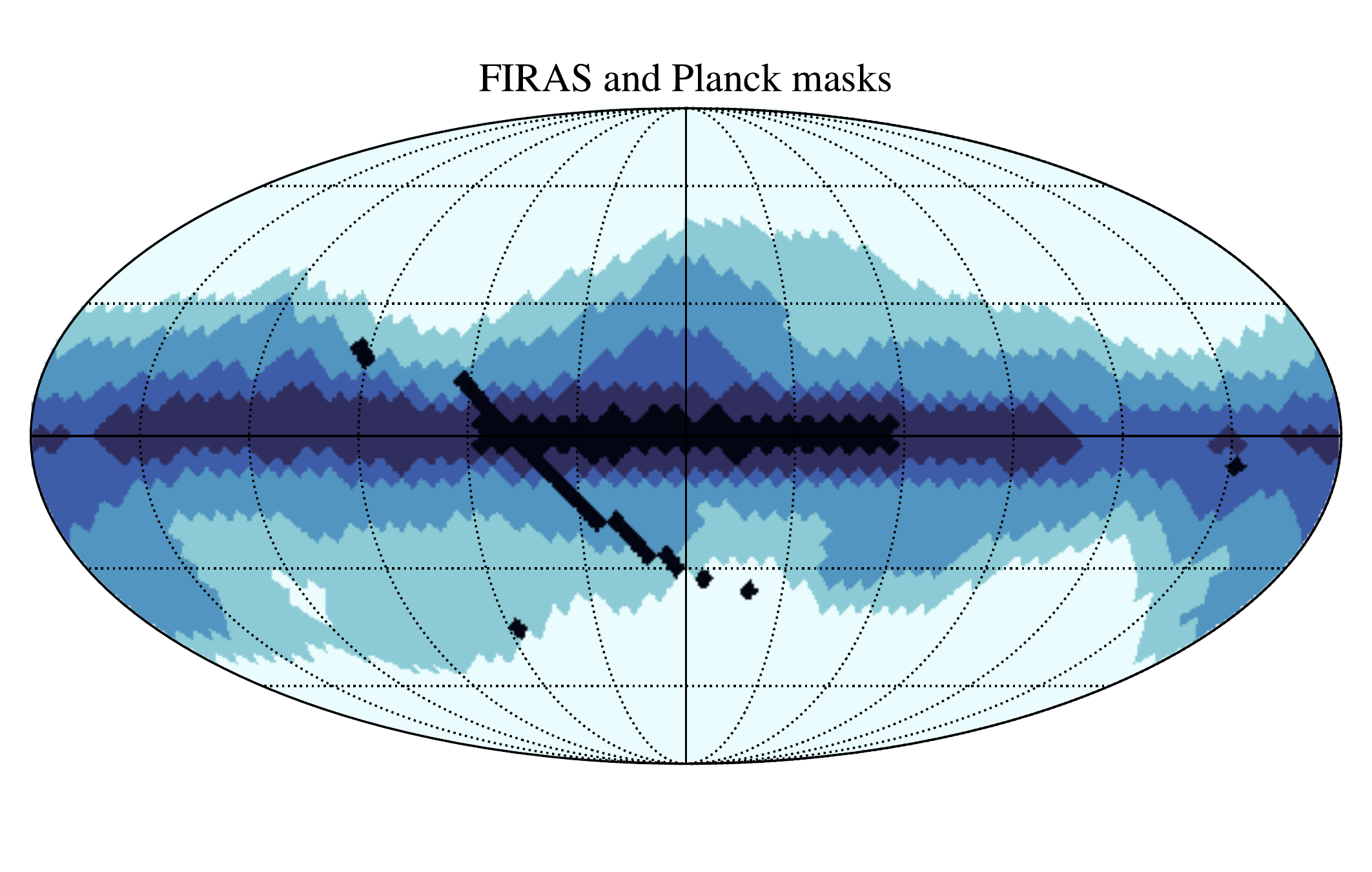}
\caption{Sky masks used in this analysis. The black region corresponds to the area masked out by the FIRAS destriper mask while progressively paler colors show the \planck masks that retain the 90\%, 80\%, 60\%, and 40\% of the sky.}
\label{fig:masks}
\end{figure}

We use different masks to remove pixels close to the Galactic plane where the contamination from foreground emission is high. We adopt the so-called destriper mask of the FIRAS data release (hereafter FDS) that removes the sky pixels not observed by FIRAS as well as those not included in the destriping operation, which are confined to the Galactic center. In addition to this mask, we use the public binary \planck Galactic masks that retain 40\%, 60\% and 80\%, 90\% of the sky and that are derived from thresholding the 353  GHz map after CMB subtraction.
Hereafter we will refer to these masks, which are shown in Fig.~\ref{fig:masks}, as P40, P60, P80, P90 respectively and will adopt P60 as our baseline mask.

\section{\label{sec:methods}Methods} 
In this section, we describe the different steps that compose the analysis methodology. We start from the modelling of the FIRAS data cube and foreground emissions; we then discuss the sky component fitting and power spectrum estimation, and finally conclude with a description of the cosmological parameter inference framework.

\subsection{\label{sec:datamodel}Data model}
The calibrated FIRAS low frequency sky spectra are a function of frequency $\nu$ and position $\nver$. 
Effectively, we can think of the data as a set of pixelized maps at different frequencies, where the value in each pixel corresponds to the absolute sky emission in MJy/sr.
We model the sky emission $I_{\nu}(\nver)$ as a superposition of several components: 
\begin{enumerate}
    \item a blackbody at $T=T_0$ describing the CMB monopole, $B_{\nu}(T)$;
    \item a CMB dipole with spectral radiance given by the derivative of the Planck function $B_{\nu}$ with respect to the temperature (assuming it is the result of a Doppler shift of the monopole induced by our peculiar motion);
    \item  selected foreground emission components, $I_{\nu}^{\rm FG}(\nver)$;
    \item additional spectral distortions, of $\mu$- or $y$-type, $I_{\nu}^{\rm SD}(\nver)$ (in our notation hereafter ${\rm SD}\in [\mu,y]$).  
\end{enumerate}
Altogether, and similarly to \cite[][hereafter \fixsenspec]{fixsen96}, the sky model in the direction $\nver$ reads
\begin{eqnarray}\label{eq:datamodel}
    I_{\nu}(\nver) &=& B_{\nu}(T_0) + \Delta T(\nver) \frac{\partial B_{\nu}}{\partial T}\bigg\rvert_{T=T_0}\nonumber\\
     & + & I_{\nu}^{\rm FG}(\nver) + I_{\nu}^{\rm SD}(\nver),
\label{eq:data-model}
\end{eqnarray}
where $I_{\nu}^{\rm FG}(\nver) $ describes the galactic and extragalactic foreground emissions and $I_{\nu}^{\rm SD}(\nver)$ the spectral distortion spectral energy distribution (SED). 

In the case of $\mu$-type distortions, we can drop the $I_{\nu}^{\rm SD}(\nver)$ term and incorporate it into the generalized Planck's law for blackbody radiation $B_{\nu}(T,\mu)$
\begin{equation}
B_{\nu}(T_0,\mu(\nver)) = \frac{2h\nu^3}{c^2} \frac{1}{e^{x+\mu(\nver)}-1},
\end{equation}
with $x = \frac{h\nu}{k_B T_0}$, $\mu(\nver)$ being the chemical potential (at a given sky location), and $T_0$ the reference CMB monopole temperature. Throughout this work we assume the best fit value of $T_0=2.7255$ \cite{fixsen2009} but we tested that our results are robust with respect to this choice. As a matter of fact, any change in the local temperature of the CMB in a direction $\nver$ results in a shift in the amplitude of the $\Delta T$ parameter. The sum of the first two terms of Eq.~\eqref{eq:data-model} represents in fact a Planck blackbody spectrum of temperature $T_0+\Delta T$. Since we do not make any assumption on its amplitude nor its spatial dependence, the $\Delta T$ parameter effectively captures changes in the CMB temperature at linear order and essentially removes residual CMB contaminations in our resulting SD estimate, allowing us to marginalize over this effect in the estimate of $\mu$. We note that including the $\mu$ distortion as a non-linear parameter in the fit, contrary to \fixsenspec, allows us to minimize degeneracies between CMB and $\mu$ (see Fig.~\ref{fig:contour_Dbeta}). In fact, considering $\mu$ as a linear deviation around a  Planck blackbody spectrum we would have
\begin{equation}
I_{\nu}^{\mu}(\nver) =  \mu(\nver) \frac{-T_0}{x}\frac{\partial B_\nu}{\partial T}.
\label{eq:mu-fixsen96}
\end{equation}
Given that in Eq.~\eqref{eq:data-model} and Eq.~\eqref{eq:mu-fixsen96} both the free parameters $\delta T$ and $\mu$ multiply a $\partial B_\nu/\partial T$ term, they become almost degenerate and display a very strong correlation (over 95\% according to \fixsenspec).

A similar approach can be followed to separate $y$ distortion map, where the emission law is
\begin{equation}
     I_{\nu}^{{\rm SD},y}=y T_0 \left(x\frac{e^x+1}{e^x-1}-4 \right) \frac{\partial B_{\nu}}{\partial T}\bigg\rvert_{T=T_0}.
\end{equation}
The $y$-type distortion has a shape that is similar to that of the $\mu$-type, but features a zero-crossing at $\nu\simeq$ 218  GHz instead of $\nu\simeq$ 125  GHz. 

Since the focus of our work is the $\mu$-type distortion and its potentiality to constrain primordial non-Gaussianities, in our baseline analysis we only consider $\mu$ as additional spectral distortion components, but in Sec.~\ref{sec:systematics} we explore how including the $y$-type distortion in the fit affects our results.

\subsection{\label{sec:fgmodels} Foreground modelling}
We discuss here the different choices regarding the parametrization of the foreground emission $I_{\nu}^{\rm FG}$ in total intensity.\\

{\bf Galactic dust}. Dust grains in the interstellar medium absorb UV light from hot stars and reradiate in the sub-millimeter and infrared bands, dominating the foreground emission at frequencies $\nu \gtrsim 100$  GHz. We model thermal emission from Galactic dust as a modified blackbody described by a dust temperature $T_d=19.6$ K and a spectral slope of $\beta=1.6$, which is scaled at a reference frequency of 353  GHz following \citep{planck15_fg}:
\begin{equation}
    I_{\nu}^{\text {dust }}(\nver) = A_d(\nver) \left(\frac{\nu}{353}\right)^{\beta_d(\nver)} \frac{B_{\nu}\left(T_{d}\right)}{B_{353 \rm  GHz}\left(T_{d}\right)}.
\label{eq:dust-emission}
\end{equation}
We infer the dust brightness in each pixel, $A_d(\nver)$ (expressed in MJy/sr).
In the following we will consider two scenarios: one where we keep the dust spectral index fixed to its fiducial value of 1.6, and one where we allow for spatial variations of $\beta_d\equiv\beta_d(\nver)$, i.e. we fit for $A_d$ and $\beta_d$ separately in each pixel. \\

{\bf Galactic synchrotron}. Relativistic cosmic-ray electrons accelerated by magnetic fields in our galaxy produce synchrotron radiation. The specific spectral energy distribution depends on the strength of the magnetic fields as well as the energy and flux of the electrons, and typically results in a power law spectrum. Synchrotron radiation represents the dominant foreground contribution for observations of the CMB at low frequencies. We include a synchrotron component modelled as a power law with $\beta_s=-3.1$ and $\nu_{\rm ref}=23$  GHz, i.e.
\begin{equation}
    I_{\nu}^{\rm sync}(\nver)= A_s(\nver) \left( \frac{\nu}{23 \rm  GHz}\right)^{\beta_s}
\end{equation}
following \cite{Planck2015_X}. When including synchrotron as a foreground component, we fit for $A_s(\nver)$ in each individual pixel.\\

{\bf Free-free}.  Thermal bremsstrahlung (free-free) emission of free electrons in star-forming regions within our Galaxy is an important foreground to CMB observations at low and intermediate frequencies. To remove this emission, which is mainly concentrated in the Galactic plane, we use the \planck Commander free-free template maps for the emission measure ${\rm EM}(\nver)$ and electronic temperature $T_{\rm e}(\nver)$ \citep{Planck2015_X}, and we follow the recipe of the same paper to rescale the intensity of the emission across frequencies so that
\begin{align}
I^{\,\rm ff}_\nu(\nver) = 10^6\, T_{\rm e}(\nver)\left(1-{\rm e}^{-\tau_{\rm ff}(\nver,\nu)}\right)\,,
\end{align}
where the free-free optical depth is given by
%-----------------------------
\begin{align}
\tau_{\rm ff}(\nver,\nu) = 0.05468\,T_{\rm e}(\nver)^{-3/2}\,\left({\nu\over 1\,{\rm  GHz}}\right)^{-2}\,{\rm EM}(\nver)\,g_{\rm ff}(\nver,\nu)\,,
\end{align}
%-----------------------------
and the Gaunt correction factor is %-----------------------------
\citep{Draine2003}
\begin{align}
g_{\rm ff}(\nver,\nu) &=1+\log\left(1+ {\rm e}^{4.960+{\sqrt{3}\over \pi}\log\left[\left({\nu\over 1\,{\rm  GHz}}\right)^{-1}\left({T_{\rm e}(\nver)\over 10^4\,{\rm K}}\right)^{3/2}\right]} \right)\,.
\end{align}
%-----------------------------
For each sky pixel we then fit an amplitude of the emission template $A_{\rm ff}(\nver)$.\\

{\bf FIRAS Galactic residual}. To complement the physically motivated foreground emission models introduced above, we also consider the Galactic residual template from \fixsenspec. This Galactic spectrum is empirically derived under the assumption that it correlates spatially with the average intensity from the high frequency channels at each pixel.\\

All foreground components, except for the Galactic thermal dust, are described by one free parameter only, namely the amplitude of the foreground template in each pixel. 
In our baseline analysis (hereafter referred to as $A_d+\beta_d$ model) we only include emission from Galactic thermal dust where both the amplitude and spectral index are allowed to spatially vary (for a total of four free parameters per pixel) but consider alternative scenarios: Galactic dust with a spectral index fixed to $\beta_d=1.6$ ($A_d$ model), Galactic dust with $\beta_d=1.6$ plus synchrotron emission ($A_d+A_s$ model), Galactic dust with $\beta_d=1.6$ plus free-free emission ($A_d+A_{\rm ff}$ model), and the FIRAS model. Finally, we note that in all the scenarios we considered in this work we neglect the molecular lines emission as well as the Cosmic Infrared Background (CIB) emission produced by unresolved dusty star forming galaxies. As far as the CIB is concerned, our baseline model could in principle not only  capture the complexity of the dust emission across the sky but also to (at least) partially account for the CIB emission itself. In fact, the CIB itself can be described by a modified blackbody SED with a different $T_d$ compared to the one of the Galactic dust. Given that $T_d$ and $\beta_d$ are usually highly correlated parameters, our data model can in principle effectively capture the emission of two superposing modified blackbody emission (i.e. Galactic dust and CIB) in a given sky pixel. More complex models should be able to capture this effect more accurately \cite{Chluba:2017rtj}. Moreover, in our analysis we do not include frequencies above 600 GHz where the CIB becomes more important outside of the Galactic plane. For molecular lines, the frequency range we consider covers the CO lines from the (J=1-0) transition at 115.57 GHz to the (J=5-4) transition at 576 GHz as well as the C-I transitions can become important. Since these line emissions are mainly concentrated in the Galactic plane, we do not include them in the analysis. We note that since line emission and $\beta_d$ are degenerate \cite[see e.g. the discussion in Sec 4.4. of][]{so-galscience}, our fitting could also help partially removing them. Being the signature of $\mu$ distortions relatively more important at lower frequencies, the channels potentially contaminated by these line emission have a lower weight in the final component separation fit due to their higher noise, and thus we do not expect them to become important.

We test the sensitivity of our results with respect to the assumed foreground model in Sec.~\ref{sec:systematics}.

\subsection{FIRAS covariance}
A key ingredient to carry out cosmological parameter inference from FIRAS data is the sky spectra covariance matrix which describes the correlation structure of observations between different frequencies and sky pixels.
As discussed extensively in the FIRAS Explanatory Supplement \cite{firas-expsup}\footnote{Available in electronic form at \url{https://lambda.gsfc.nasa.gov/product/cobe/firas_exsupv4.html}}, there are six main sources of uncertainties: detector noise (C vectors, using the FIRAS naming), bolometer parameter gain uncertainties (JCJ), emissivity gain uncertainties (PEP), internal calibrator temperature errors (PUP), absolute temperature errors (PTP), and destriper errors $\beta$ (which include map offsets uncertainties).
Note that some of these quantities only vary across pixels, such as $\beta$, while others only differ between frequencies (PEP, JCJ, PUP and PTP).

\begin{figure}[t]
    \centering
    \includegraphics[width=\columnwidth]{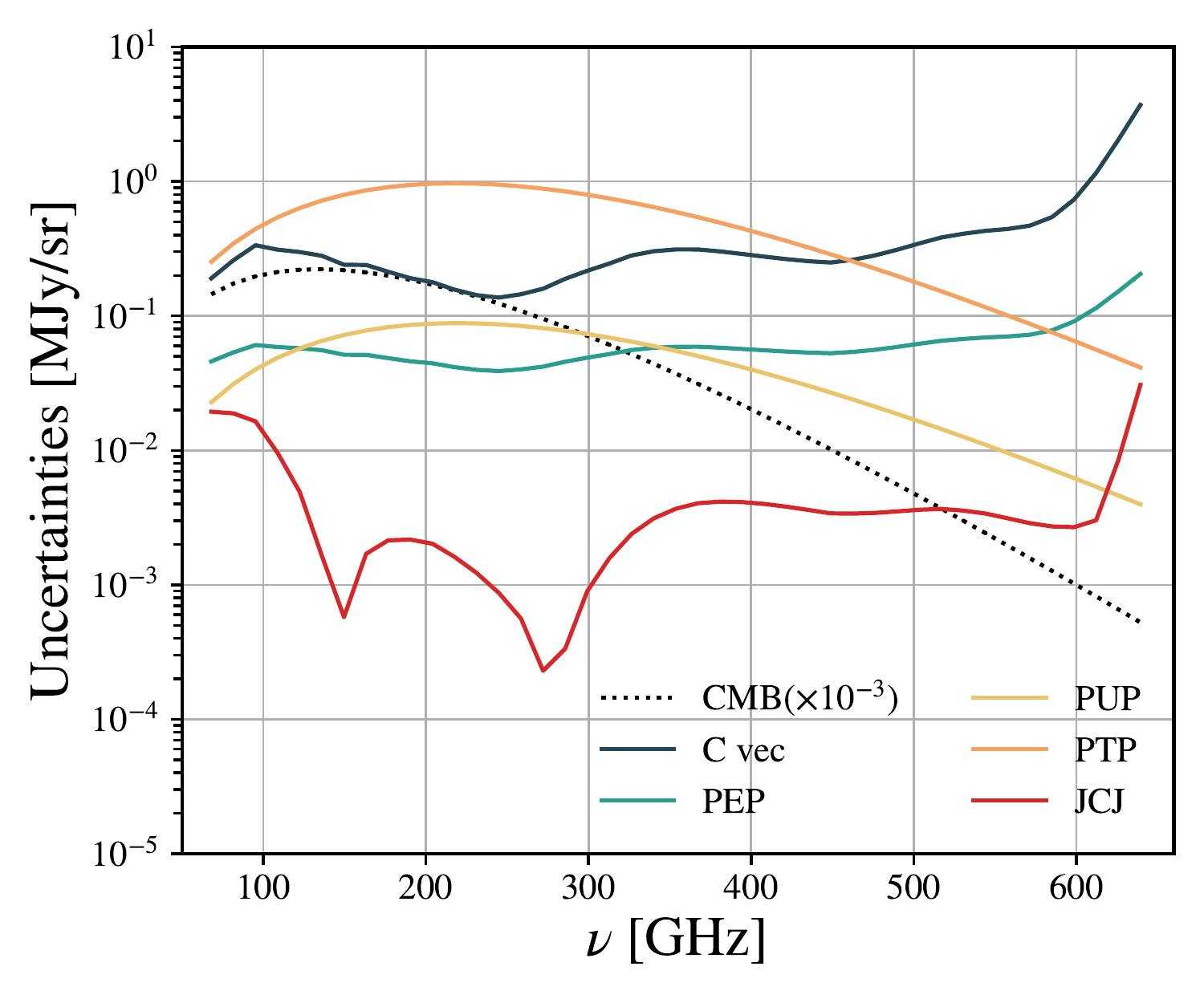}
    \caption{Noise budget of the FIRAS low frequency low spectral resolution Destriped Sky Spectra. The dotted black line shows the CMB blackbody emission rescaled by a factor $0.001$ for visualization purposes. For the JCJ and PEP errors described in Eq.~\eqref{eq:gain-errors} which depend on the sky pixel, we display their average value computed across all the observed pixels.}
    \label{fig:firas_errors}
\end{figure}

Denoting an element of the FIRAS data cube $\hat{I}_{\nu}(\nver)\equiv \hat{I}_{\nu p}$, we assemble the FIRAS covariance matrix as
\begin{eqnarray}
    \mathbb{C}_{\nu p\nu'p'} &=& {\rm Cov}(\hat{I}^{\rm FIRAS}_{\nu p},\hat{I}^{\rm FIRAS}_{\nu'p'})\nonumber \label{eq:full-cov}\\
    &=& C^{\nu \nu^{\prime}}\left(\delta^{p p^{\prime}} / N_{p}+\beta_{k}^{p} \beta_{p^{\prime} k}+0.04^{2}\right) \label{eq:pixel-cov} \\
    &+& S^{p \nu} S^{p^{\prime} \nu^{\prime}}\left(J^{\nu} J^{\nu^{\prime}}+G^{\nu} G^{\nu} \delta^{\nu \nu^{\prime}}\right) \label{eq:gain-errors}\\
    &+& P^{\nu} P^{\nu^{\prime}}\left(U^{2} \delta^{p p^{\prime}} / N_{p}+T^{2}\right).
\label{eq:fullcov}
\end{eqnarray}
In the equation above, $N_p$ is the map pixel weight, $\beta_k^p$ is the $\beta$ matrix described in Section 7.2.2 of the FIRAS explanatory supplement and we sum $\beta^p_k\beta_{pk^\prime}$ over all $k$-th orthogonalized stripes used in the data destriping. The $S^{p\nu}$  vectors represent the absolute sky brightness not including the CMB monopole and $J^\nu$ is the JCJ gain term described in Section 7.3.2 of the FIRAS explanatory supplement while the $G^\nu$ terms describe the PEP gain errors. We use the FIRAS data measurement themselves $\hat{I}_{\nu p }$ from which we subtract the CMB monopole adopting a temperature of  $T_0=2.7255$. We assume  $P^\nu = \partial B_\nu(2.728 K)/\partial T$ and $U$, i.e. the internal calibrator temperature uncertainty, is 180 $\mu$K as suggested in Section 7.4.5 of the FIRAS explanatory supplement. As for the PEP and JCJ terms, we take the PTP terms from the publicly available calibration uncertainty file available on LAMBDA so that $T=0.002$ K. 
In Fig.~\ref{fig:firas_errors}, we show a breakdown of the different uncertainty terms as function of frequency $\nu$, noting that the main sources of error over the range of frequencies considered here are the absolute temperature errors and detector noise.

\subsection{\label{sec:mapfit} Map inference}
With a data model and an estimate of the measurement uncertainties at hand, we can now reconstruct the spatial fluctuations of the effective chemical potential $\mu(\nver)$ as well as the other parameters $\bm \theta$ that describe the data model.

In the following we will assume that the FIRAS noise covariance is approximately uncorrelated between pixels while still retaining the full frequency-frequency structure.  We investigate the validity of this assumption and its impact on the angular power spectrum estimation level in Sec.~\ref{sec:ps_ext}.
Assuming pixels are uncorrelated greatly simplifies the analysis since it allows us to perform the parameter inference at the individual pixel level.
To this end, we use Bayes theorem and write the posterior distribution for the model parameters in each pixel as $\propto \ln\mathcal{L}(\hat{I}_{\nu}|{\bm\theta})p(\bm\theta)$, where the likelihood function is given by\footnote{Omitting the pixel dependence and up to a normalizing constant.}
\begin{equation}\label{eq:pix_like}
    -2\ln\mathcal{L}(\hat{I}_{\nu}|\theta) = \sum_{\nu\nu'}\Delta_{\nu}^{T}(\bm\theta) \mathbb{C}^{-1}_{\nu\nu'} \Delta_{\nu'}(\bm\theta),
\end{equation}
with $\Delta_{\nu}(\bm\theta)$ denoting the residuals between the observed FIRAS spectra and the model spectrum $\Delta_{\nu}(\mathbf{\theta}) = \hat{I}_{\nu}-I_{\nu}(\mathbf{\theta})$, and $\mathbb{C}^{-1}_{\nu\nu'}$ being the inverse of the covariance matrix defined in Eq.~\eqref{eq:fullcov} for $p=p^\prime$.
We use \texttt{emcee}, an affine-invariant Markov Chain Monte Carlo (MCMC) sampler \citep{emcee}, to explore the posterior distribution and use $N= 100$ walkers for 1000 steps to reach the convergence of the chains (assessed by checking their auto-correlation time).
This approach allows us to estimate the full non-Gaussian posterior, the level of degeneracy between different parameters, and to directly marginalize over the impact of foregrounds.
In our analysis we do not impose priors on the sampled parameters except for the spectral slope of the Galactic thermal dust, for which we set a uniform prior over $\beta_d \in [0,3]$.

\begin{figure}[t]
    \centering
    \includegraphics[width=\columnwidth]{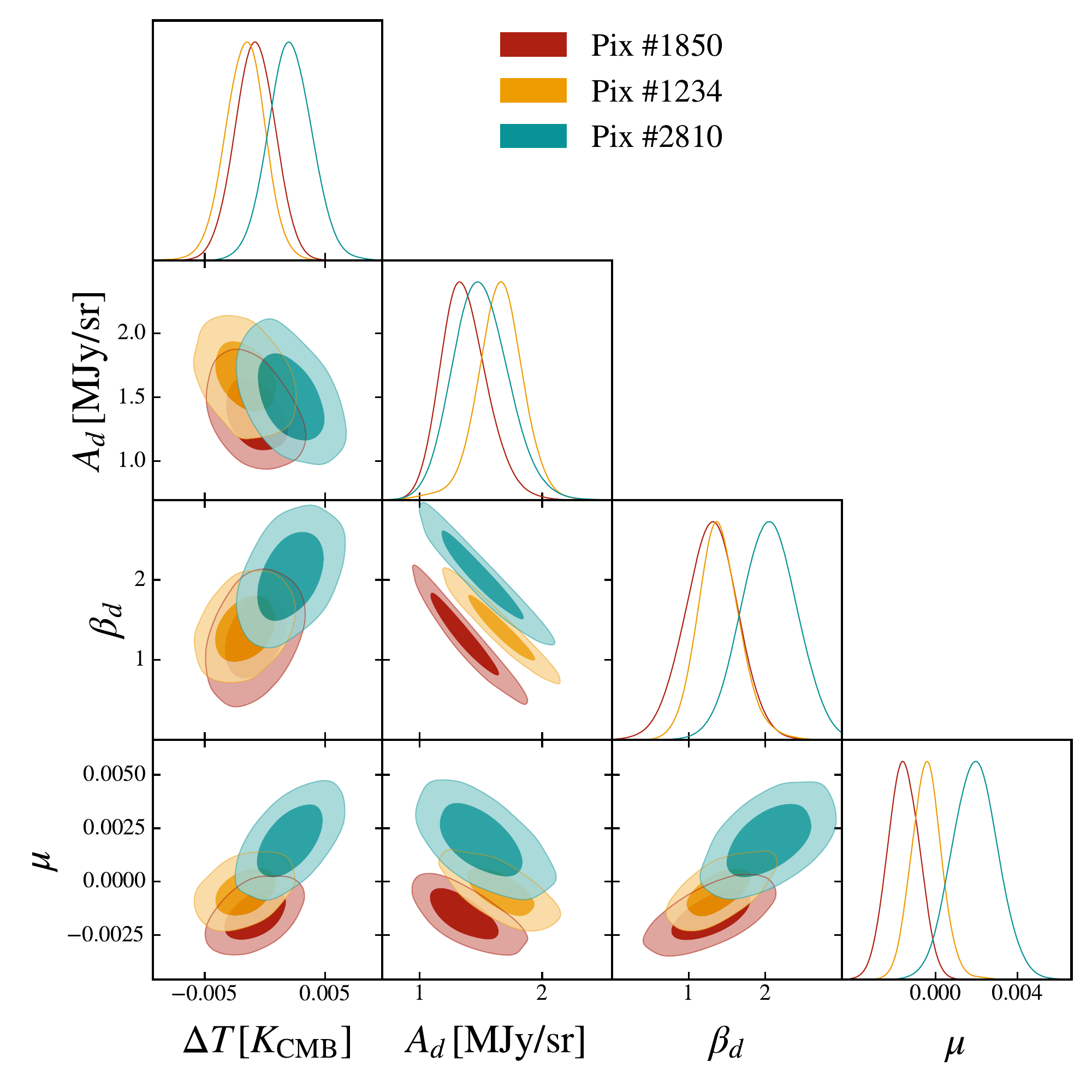}
    \caption{Constraints on the baseline data model parameters from FIRAS sky spectra. Different colors show results for three different pixels. The darker and lighter shaded regions represent the 1 and 2$\sigma$ contours respectively. Typically, our fitting also reduces the correlation between $\mu$ and $\Delta T$ from the 95\% level observed in previous analyses.}
    \label{fig:contour_Dbeta}
\end{figure}

In Fig.~\ref{fig:contour_Dbeta}, we show a representative set of credibility contours in the $\{\Delta T,A_d,\beta_d,\mu\}$ parameter space from our baseline data model (Galactic dust with free spectral index) for three different pixels.
We can note an anti-correlation between the amplitude and spectral index of the Galactic dust and a positive correlation between $\mu$ and the CMB temperature.
The extent of the two-dimensional contours reflects the level of noise in a given pixel.
The one-dimensional $\mu$ posteriors marginalized over the remaining parameters generally follow a Gaussian distribution.

The result of this process is a set of $\mu$ values for each pixel in an \textsc{HEALPix} map at $N_{\rm side}=16$, along with an estimate of their uncertainty calculated as the standard deviation of the $\mu$ posterior.

\subsection{\label{sec:ps_ext}Power spectrum and uncertainties estimation}
After reconstructing the $\mu$ distortion anisotropy map from the FIRAS data cube, the next step consists in extracting its angular cross-power spectra with maps of primary CMB anisotropies from \textit{Planck}. 

A great deal of the information encoded in the SD-CMB cross-spectra resides in the lowest multipoles \citep{pajer12,ganc12}, therefore it is crucial to build an analysis pipeline that optimally recovers the information on the largest angular scales.
For this reason, and given the fact that we work with a dataset a at coarse \textsc{HEALPix} resolution, we use a pixel-based Quadratic Maximum Likelihood (QML) power spectrum estimator \xqml \cite{vanneste18}, an extension of the original method introduced in \citet{tegmark97} to cross power spectra. Such methods are computationally more expensive compared to traditional pseudo-$C_{\ell}$ methods but we find that they produce measurement of the cross-correlation power spectra with roughly twice smaller error bars for the multipoles most relevant for our analysis. The application of a QML power spectrum estimator requires the knowledge of the covariance of the maps as well as the details of the optical response of the instrument. Below we detail the analysis choices we made. 

\subsubsection{FIRAS beam transfer function and deconvolution}
We compute the FIRAS beam response by performing an Hankel transform of the publicly available FIRAS instantaneous beam radial profile. 
In addition to this smoothing, it is important to take into account the instrument scan motion during the integration of a FIRAS interferogram. 
The telescope motion causes the maps to be additionally smoothed in the ecliptic scan direction. 
We account for this effect by deconvolving an effective transfer function estimated from simulations as described in \cite[][]{odegard2019,anderson22}.
In this approach, high resolution realizations of CMB maps are first smoothed with the instantaneous FIRAS beam and then smoothed in real space applying a $2.4$\textdegree boxcar average in the ecliptic direction. The transfer function is then computed by comparing the power spectrum of these maps with the one of the theoretical model used to generate the input CMB maps. We multiply the pixel window function of the \textsc{HEALPix} maps by this transfer function to obtain the total transfer function of the maps that we apply to both the CMB and $\mu$ harmonic coefficients when computing auto and cross power spectra with \xqml . 

\subsubsection{Covariance matrix of CMB maps}
The noise of the \planck component-separated CMB maps can be modelled in its full complexity using the publicly available FFP10 simulations which include not only the detector noise in the time-ordered data (TOD), including its correlated $1/f$ part, but also realistic simulations of instrumental effects for all \planck frequency channels. 
These simulated TOD are then processed with the same algorithms as for the flight data, including component separation.  
We use the data available at NERSC supercomputing center that include 300 realizations of noise and residuals systematic effects for all the component-separated CMB maps and for different data splits (full- and half-mission). 
After accounting for the \planck beam smoothing, we apply the same harmonic domain transfer function that we apply to the data to all 300 realizations for all splits, and downgrade each realization to $N_{side}=16$ \textsc{HEALPix} resolution. 
Given that only 300 noise realizations are available, estimating a full dense pixel-pixel noise covariance matrix for the full mission dataset is non-trivial.
The reason is that the number of matrix elements to be estimated is much larger than 300, making the estimate of the inverse of the covariance matrix poorly conditioned.
We therefore consider only a diagonal covariance matrix where the diagonal is given by the variance of the 300 noise realizations for all the $T, Q, U$ Stokes parameters. 
While an optimal analysis of the low multipole CMB signal would require an estimate of the off-diagonal terms, for cross-correlations between independent datasets we expect these correlations to be of marginal importance. We also verify that the diagonal term of the $QU$ block of the noise covariance is negligible and we therefore discard it in our analysis. 
We perform the same operation on the half-mission splits in order to produce a noise model to be employed for null tests analysis and consistency checks. 
The signal covariance is instead computed on the fly while estimating the power spectrum. 
To this end, we adopt the fiducial lensed CMB power spectrum of the cosmology from the FFP10 simulations. 
This assumes a \planck 2018 cosmology with a tensor-to-scalar ratio $r=0$ with the CMB dipole and monopole removed. We stress that an accurate analysis, in particular of CMB polarization, requires the use of the realistic FFP10 simulations since naive estimates of the noise levels relying on the assumption of a Gaussian noise drawn from the noise covariance matrix of the frequency maps largely underestimate the uncertainties on the scales considered in this work.

\subsubsection{Covariance matrix of $\mu$ map}\label{sec:mucov}
\begin{figure}
\includegraphics[width=.9\columnwidth]{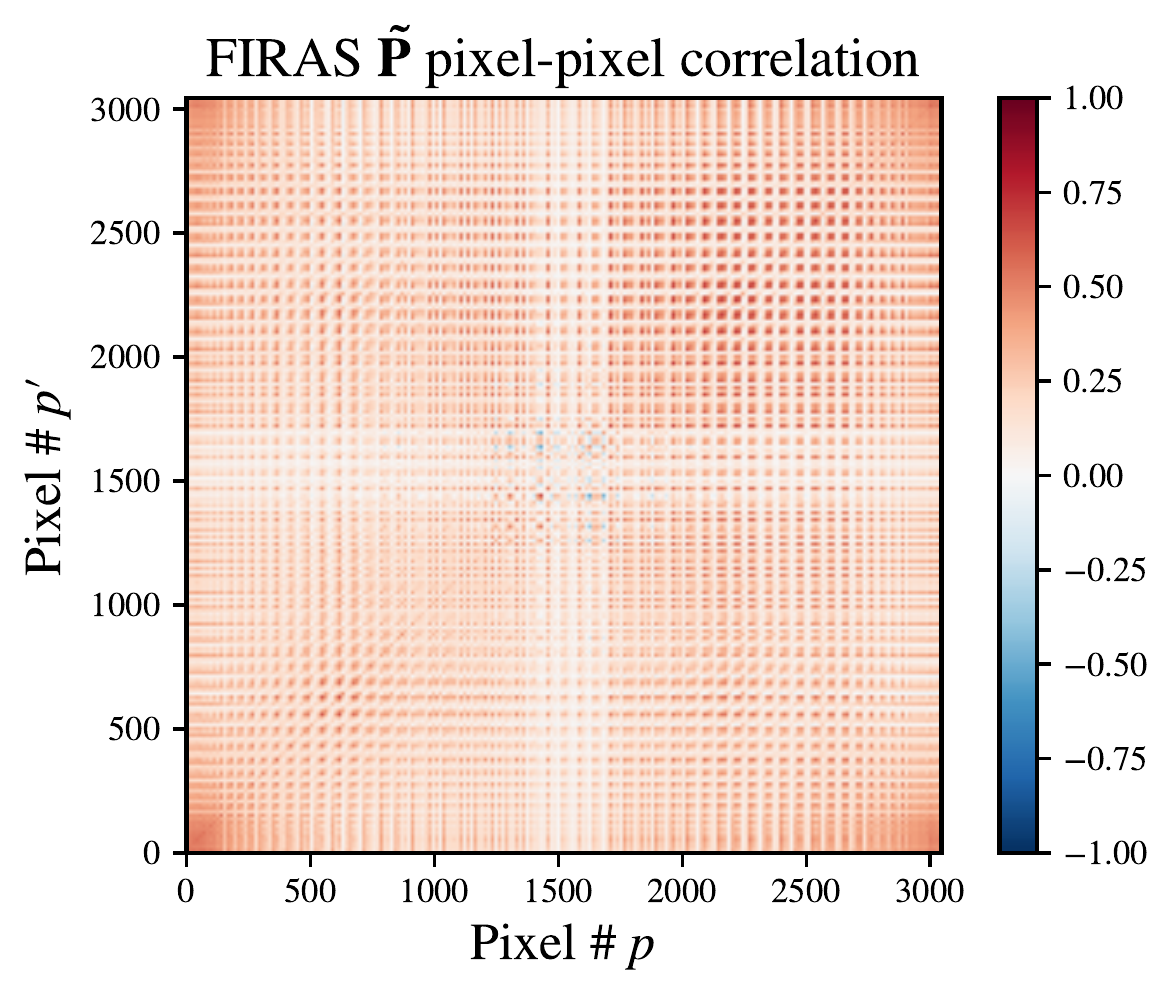}
\caption{FIRAS pixel-pixel correlation matrix. The pixels are ordered in the ring scheme of \textsc{Healpix} pixelization, i.e. pixel 0 correspond approximately to the north Galactic pole and the last one to the South Galactic pole. The median correlation coefficient across all measured pixel is about 10\% but increase to 20\% for pixels far from the Galactic plane. }
\label{fig:pixpix-corr}
\end{figure}

\begin{figure}
\includegraphics[width=\columnwidth]{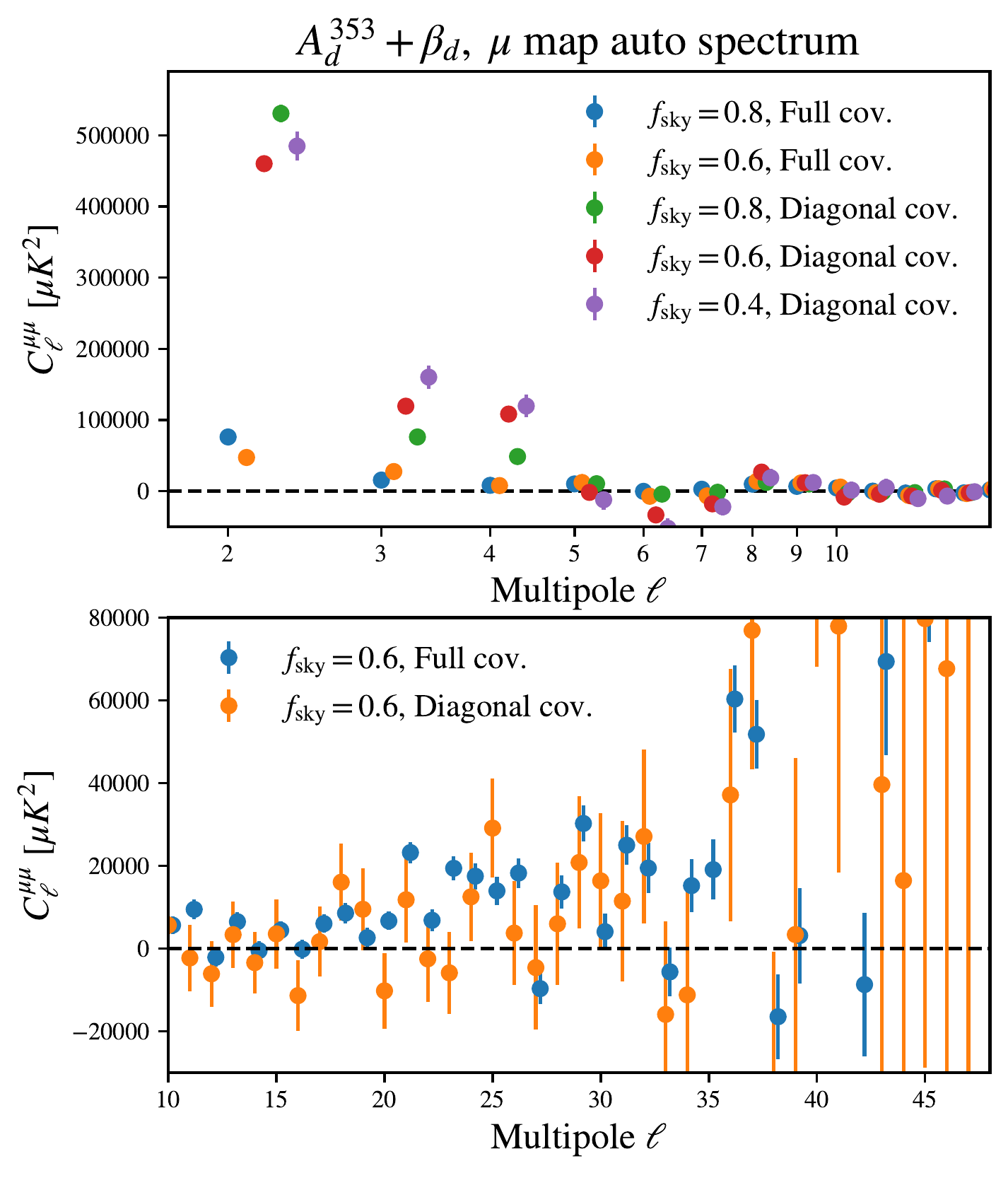}\\
\includegraphics[width=\columnwidth]{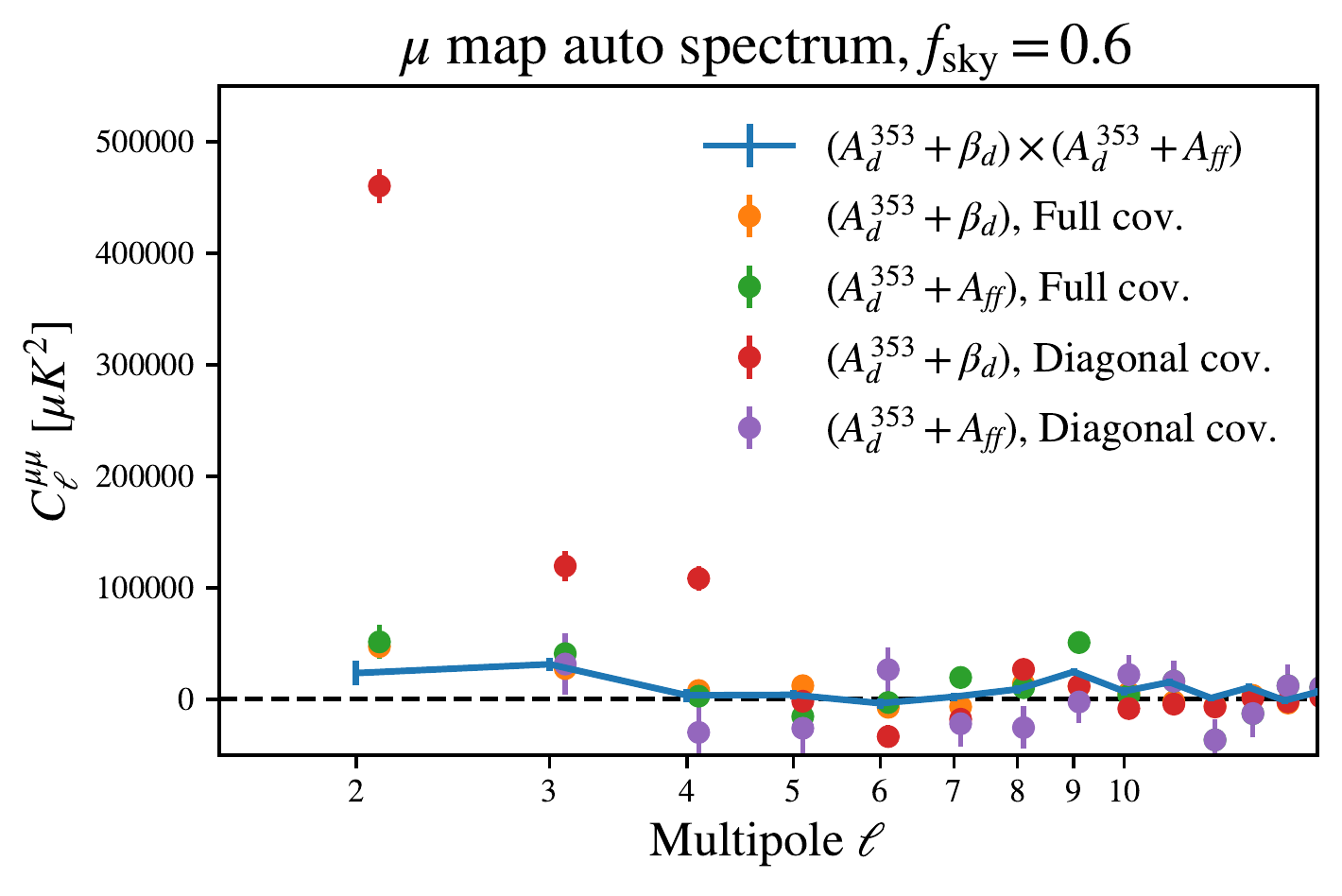}
\caption{\textit{Top}: noise-debiased auto-spectrum of our $\mu$ baseline map on large angular scales for different Galactic masks and different approximation to the pixel-pixel covariance of the $\mu$ map. The diagonal covariance does not describe the data accurately enough while the full covariance model provides a more accurate description of the data. Middle: same as above for angular scales $\ell>10$. In this regime the full covariance model for the $\mu$ map is inaccurate and only the diagonal approximation leads to a noise-debiased $\clmumu$ consistent with the null hypothesis. \textit{Bottom}: same as top panel for $\mu$ maps obtained with different component separation methods. Different $\mu$ maps have different noise properties and their cross-correlation reduces the residual noise bias in the autospectrum. We show an example of this in solid blue.}
\label{fig:mumu}
\end{figure}

When estimating the $\mu$ map, we assume that all pixels are uncorrelated between each other and, under this assumption,  for each pixel we obtain an estimate of the error on the inferred $\mu$ value in the $p$-th pixel, $\sigma^\mu_p$, from the standard deviation of the MCMC posterior. 
However, the covariance definition in Eq.~\eqref{eq:full-cov} shows a non-diagonal structure in pixel space, thus the estimate of the $\mu$ signal in each pixel should show a degree of correlation between pixels similar to the one of  the FIRAS data themselves. Therefore, we assumed the $\mu$ map is composed by noise only and we adopted two methods to compute its noise covariance matrix:

\begin{itemize}
\item We assume the $\mu$ map covariance to retain a diagonal structure $\mathbf{C}^{\mu}_{pp^\prime} = \sigma^\mu_p\delta_{pp^\prime}$. We refer to this approximation as diagonal covariance in the following. 
\item We assume the dominant off-diagonal component of the FIRAS pixel-pixel correlation matrix to be inherited by the $\mu$ map. Defining the matrix $\mathbf{P}$ from Eq.~\eqref{eq:pixel-cov}  as
\begin{equation}
P_{pp^\prime} \equiv \delta^{p p^{\prime}} / N_{p}+\beta_{k}^{p} \beta_{p^{\prime} k}+0.04^{2},
\end{equation}
we normalize $\mathbf{P}$ to obtain a correlation matrix $\mathbf{\tilde{P}}$ that we show in Fig.~\ref{fig:pixpix-corr} for reference. 
We then convert $\mathbf{\tilde{P}}$ into a dense pixel-pixel covariance matrix for the $\mu$ map $\mathbf{C}^\mu_{pp^\prime}$ using the error estimate from the MCMC as 
\begin{equation}
\mathbf{C}^\mu_{pp^\prime}  = \frac{P_{pp^{\prime}}}{\sqrt{\text{diag}(\mathbf{P})_p \text{diag}(\mathbf{P})_{p^{\prime}}}}\cdot\sqrt{\sigma^{\mu}_{p}\sigma^{\mu}_{p^{\prime}}}.
\end{equation}
\end{itemize} 
For each different $\mu$ map derived with a different component separation method, we compute the corresponding covariance matrix and use that in the power spectrum estimation step.
These two methods describe the noise properties of the maps with different accuracy in different regimes. 
For FIRAS, no data split that can be used to validate the noise covariance matrix through, e.g., a jackknife approach or a null map is available. 
As such, we validate the noise model by debiasing the autospectrum of the $\mu$ maps and checking its consistency with the null hypothesis. 
For this purpose, we generate random realizations of the $\mu$ map starting from each different covariance model and multiplying the Cholesky decomposition of $\mathbf{C}^\mu$ by a stream of Gaussian pseudo-random numbers with zero mean and unit variance. 
We then compute the noise bias as the mean of the simulations and the error bar as the standard deviation of the same set. 
The results of this test for different galactic masks are shown in Fig.~\ref{fig:mumu}. 
As we can see, the diagonal $\mu$ covariance does not describe the data at large angular scales where we observe a large excess of power that is however greatly reduced if the full covariance model is adopted. The opposite applies when considering angular scales $\ell\gtrsim 10$, where the full covariance model does not deliver an angular power spectrum consistent with zero. 
In both cases, the $\chi^2$ test yields a low value of the PTEs and a single covariance model is not capable to describe the data on all angular scales. 
As such, in the following we decide to adopt a hybrid approach for the power spectrum estimation and computed cross-spectra using the full covariance matrix for $\mu$ for multipoles $\ell<10$, while the diagonal covariance is adopted for $\ell\geq 10$. 
The cut at $\ell=10$ was defined as it is the lowest multipole for which the $\chi^2$ test for consistency with the null hypothesis for $\clmumu$ leads to a PTE higher than 5\% for the diagonal covariance model. 
At large angular scales, the same $\chi^2$ test always leads to a PTE lower than 5\%. 
However, the $\chi^2$ value for the full covariance model is 10 times lower than the one obtained for the diagonal covariance approximation and is driven by the high value of $\ell\leq 4$, pointing to a more accurate description of the data. 
We will see that this approximation and hybrid approach to power spectrum estimation of the data is accurate enough to describe the cross-correlation power spectra between $\mu$ and the CMB.  
Finally, we note that an excess of power at large scales might be due to foreground residuals. 
We test this hypothesis by checking the stability of our estimate as a function of sky fraction used for $\clmumu$ estimation. 
As we can see in Fig.~\ref{fig:mumu}, using a mask covering progressively smaller sky fractions gives to an increased power, which is inconsistent with the expectation of the excess of power being due to foreground contamination as we would expect it to decrease with decreasing sky fraction. 
As such we conclude that inaccurate noise bias effects dominate the estimate of $\clmumu$. 
This conclusion is supported also by the fact that if we take the cross-spectra between $\mu$ maps obtained with different component separation methods, the noise is only partially correlated between the two maps as the data model that is fitted to the data is different. As such, it is partially free from residual noise bias. 
In the bottom panel of Fig.~\ref{fig:mumu} we show the bandpowers obtained by cross-correlating the $\mu$ maps obtained with the baseline and the $A_d+A_{\rm ff}$ model adopting our hybrid power spectrum estimation approach, and compare them with the debiased auto-spectra. 
As we will discuss further below, $A_{\rm ff}$ does not detect any significant foreground power and as such, the reduced amplitude of the power in the first few multipoles is caused by a lower residual noise bias rather than by an improved foreground subtraction. 

\subsection{$\clmut, \clmue, C_{\ell}^{\mu B}$ estimation}
When correlating the $\mu$ map with CMB temperature and polarization maps, we adopt the hybrid approach to evaluate the largest scales outlined in the previous section.
In doing so, we also marginalize over spurious monopole and dipole coupling induced by the galactic masking by summing a component proportional to the $\ell=1$ Legendre polynomial (i.e. $P_1\propto \cos(\theta_{ij})$, where $\theta_{ij}$ is the scalar product between the direction of the $i$-th and $j$-th pixels of the map) and having an amplitude of 1 $K^2$. 
We estimate the error bars on the band powers through Monte Carlo (MC) realizations of the CMB and $\mu$ distortion signal. 
For the CMB signal, we draw random realizations of the $T, Q, U$ Stokes parameters from the theoretical power spectrum of the FFP10 simulations and add the FFP10 official noise realization released by the \planck collaboration. 
We correlate each of these CMB maps with random realizations of the $\mu$ map drawn from either the full or diagonal $\mu$ covariance and take the multipoles where each of these approximations to the $\mu$ covariance becomes accurate.
We use these sets of simulations to verify that the value of the measured $\clmut$, $\clmue$, and $C_{\ell}^{\mu B}$ are consistent with the distribution of the multipoles obtained in the MC simulations.

\subsection{Cosmological inference $\fnl$}\label{sec:fNL_fit} 
The final step of the analysis pipeline is the cosmological inference. 
Specifically, we are interested in converting the reconstructed SD-CMB cross-power spectra into constraints on the amplitude of the local-type primordial non-Gaussianities, $\fnl$.
Assuming that the cross-spectra bandpowers are Gaussianly distributed, we can write
\begin{equation}\label{eq:spectra_like}
\begin{aligned}
    -2\ln\mathcal{L}(\fnl&|\hat{C}^{\mu X}_{\ell}) =\\ 
    &\sum_{\ell\ell'} \left[\hat{C}^{\mu X}_{\ell}  - \fnl C_{\ell}^{\mu X} \right] \Sigma_{\ell\ell'}^{-1} \left[\hat{C}^{\mu X}_{\ell'} -\fnl C_{\ell'}^{\mu X} \right], 
\end{aligned}
\end{equation}
where $\hat{C}^{\mu X}_{\ell}$ can either be the measured $\mu T$, $\mu E$ or $\mu B$ spectra, $C_{\ell}^{\mu X}$ are the corresponding theoretical templates calculated for $\fnl=1$, and $\Sigma_{\ell\ell'}^{-1}$ is the inverse of the bandpower-bandpower covariance matrix that we estimated from simulations and includes the effect of correlations between multipoles and correlation between the probes. We verified that the Gaussianity assumption, usually inaccurate for CMB analysis \cite{percival2006}, holds for the low, noise-dominated multipoles considered in our work computing the distribution of each of the estimated bandpowers from our simulated dataset and checking their consistency with a Gaussian distribution through a Kolmogorov-Smirnov test. For all the multipoles we indeed obtain $p$-values above 5\%. 
We account for the finite number of simulations used to estimate the bandpower covariance matrix by debiasing  $\Sigma_{\ell\ell'}^{-1}$ following the prescription from \citet{hartlap06}.
The calculation of the theoretical template spectra is detailed in Sec.~\ref{sec:fNL_constaints}.

From Eq.~\eqref{eq:spectra_like}, we can either obtain the maximum-likelihood estimate of $\fnl$ and its associated $1\sigma$ uncertainty as 
\begin{equation}
    \hat{f}_{\rm NL} = \frac{\hat{C}^{\mu X}_{\ell}\Sigma_{\ell\ell'}^{-1}C^{\mu X}_{\ell}}{C^{\mu X}_{\ell}\Sigma_{\ell\ell'}^{-1}C^{\mu X}_{\ell'}} \quad \sigma(\hat{f}_{\rm NL}) = \frac{1}{C^{\mu X}_{\ell}\Sigma_{\ell\ell'}^{-1}C^{\mu X}_{\ell'}},
\end{equation}
where we recall that the templates are calculated for $\fnl=1$, or directly sample the posterior assuming an unbound uniform prior on $\fnl$.
In this work, we present results based on both approaches.
Furthermore, we can jointly analyze the $\mu T$ and $\mu E$ power spectra by concatenating both the templates and extracted spectra into two vectors, $\mathbf{C}_{\ell}=\left[C_{\ell}^{\mu T}(\fnl=1),C_{\ell}^{\mu E}(\fnl=1) \right]^T$ and $\mathbf{\hat{C}}_{\ell}=\left[\hat{C}_{\ell}^{\mu T},\hat{C}_{\ell}^{\mu E} \right]^T$.

\section{\label{sec:results}Measurements} 
\subsection{\label{sec:maps}Spectral anisotropies, CMB and foreground maps}
\begin{figure*}[t]
    \centering
    \includegraphics[width=\textwidth]{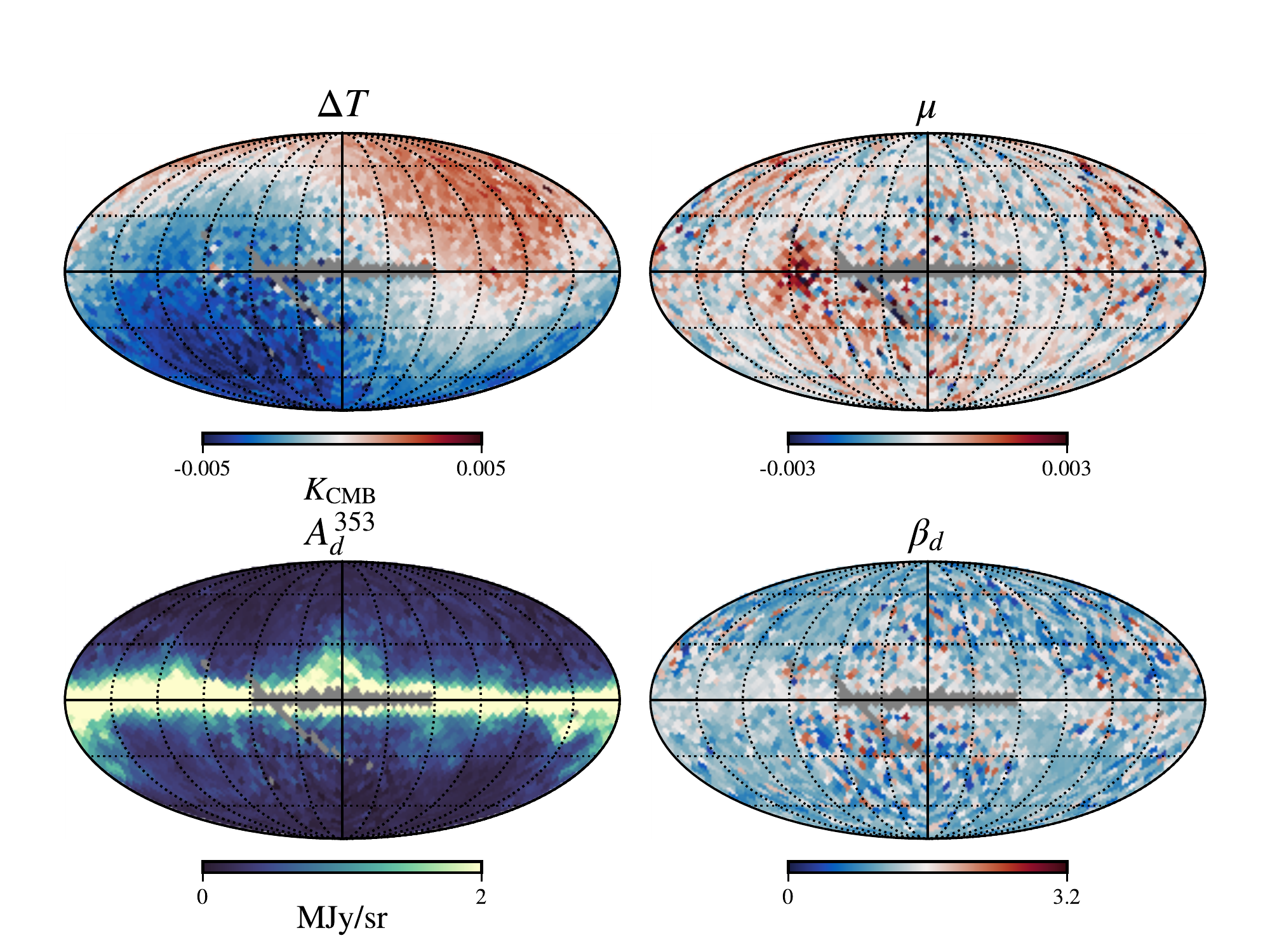}
    \caption{Maps of the CMB dipole (top left), anisotropic $\mu$-type distortion (top right), amplitude and spectral index (at 353  GHz) of the Galactic dust (bottom left and right respectively) inferred from FIRAS low frequency low spectral resolution destriped sky spectra. Pixels greyed out are removed by the FIRAS destriper mask.}
    \label{fig:maps}
\end{figure*}

We start by showing in Fig.~\ref{fig:maps} the full-sky maps of our baseline data model parameters inferred from the FIRAS data. 

\begin{figure}[!hbp]
    \centering
    \includegraphics[width=\columnwidth]{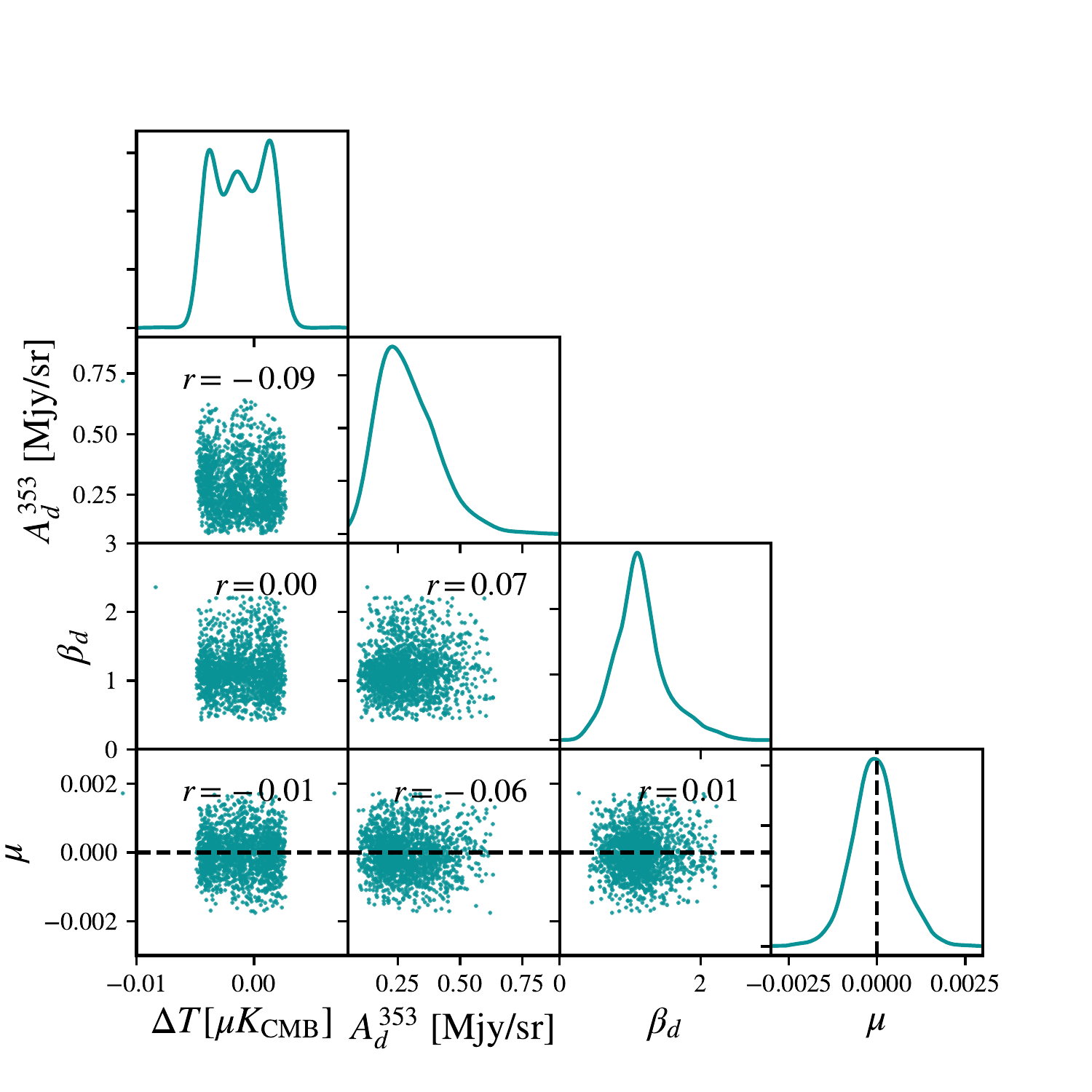}
    \caption{Scatter plot matrix of the map pixel values after applying the \planck $f_{\rm sky}=0.6$ Galactic mask. Plots on the diagonal show a Gaussian kernel density estimate of the pixel value distributions in each map. The numbers in the boxes show the Spearman's rank correlation coefficient between the two variables under consideration.}
    \label{fig:corr_Dbeta}
\end{figure}

In the top left panel we plot the CMB anisotropy map, $\Delta T$, associated with the component in the data cube that emits as $I_{\nu}\propto \partial B_{\nu}/\partial T$. 
The map has units of thermodynamic temperature $K_{\rm CMB}$ and is dominated by spatial variations due to the distinct kinematic dipole as expected. 
The top right panel shows instead the recovered map of the $\mu$-type distortion fluctuations, which is one of the main results of this paper. 
As can be seen, the map exhibits large-scale fluctuations due to the spatially-varying noise properties, while the distribution of the pixel values is centered around zero (see the lower right panel of Fig.~\ref{fig:corr_Dbeta}).
Finally, the two bottom panels show the maps that describe our baseline foreground model, the amplitude and spectral index of the Galactic thermal dust evaluated at a reference frequency of 353  GHz. 
The amplitude map $A_{d}$, in units of MJy/sr, is recovered at a high $S/N$ and faithfully traces the emission from the Galaxy.
The spectral index map is instead noisier, also showing the large-scale noise fluctuations as in the $\mu$ map case.
The color scale of the $\beta_d$ map is chosen in a way to show deviations from the reference value of $\beta_d=1.6$ \citep{BK18}.
We recall that in this work we only use the low frequency data from FIRAS since we are mainly interested in marginalizing over foreground contamination rather than providing their full characterization. 
For this specific task, the high frequency data would provide additional information.
The corresponding uncertainties on the recovered maps are shown in Appendix~\ref{sec:maps_std}.

In Fig.~\ref{fig:corr_Dbeta}, we show a scatter plot of the pixel values for each of the six different map pairs that can be constructed from the parameters in our baseline data model. 
For this check we apply our nominal fiducial P60 mask before creating the scatter plot.
The inferred pixel values are largely uncorrelated between different maps, as quantified by the Spearman's rank correlation coefficient.

The products of our analysis allow us to provide new estimates of the  $\mu$ distortion monopole, of the CMB temperature monopole $T_0$, and of the monopole of the dust emission together with its spectral index. 
We perform an inverse covariance weighted average of the map pixels and show the results of these estimates in Table~\ref{tab:mumonopole} for different Galactic masks. 
Our improved foreground modelling allows us to use a larger sky fraction compared to the original FIRAS analysis\footnote{97\% of the FDS mask compared to the fiducial one of \fixsenspec\ that retained 90\% of the sky removing pixels at $|b|<5^\circ$} and sharpen the upper limit on the monopole provided by the original FIRAS analysis $\langle \mu\rangle = (-10 \pm 40) \times 10^{-6}$ \cite{fixsen96} by roughly a factor 2 for our fiducial analysis mask $|\langle\mu\rangle| \lesssim 47\times 10^{-6}$. Constraints on models predicting energy injections from particle decays or other sources in the $\mu$ distortion era should be revised accordingly \cite[e.g.][]{Poulin:2016anj,Bolliet:2020ofj}. We obtain stable results using more aggressive Galactic masks. Adopting the same foreground model of the FIRAS analysis and removing a similar sky fraction, we recover an upper limit consistent with their original estimate of $\langle\mu\rangle \lesssim 90\times 10^{-6}$. 

\begin{table}[t]
\begin{tabular}{ccccccc}
\toprule\toprule
 $[\times 10^{-6}]$        & FDS           & PL90& PL80& PL60   & PL40  \\ 
 \midrule
$|\langle\mu\rangle| <95$\% C.L.   & $<45$  & $<47$  &$<51$ & $< 47$ & $< 53$ \\
$\langle\mu\rangle$                        & $-16.2$& $-18.2$&-21.7& $-15.5$ & $-15.6$ \\
$\sigma_{\langle\mu\rangle}$         & $14.2$ & $14.3$ &14.5 &$15.9$ & $18.6$ \\
\bottomrule\bottomrule
\end{tabular}
\caption{Statistics of the $\mu$ distortion monopole (95\% confidence level upper limits, mean and error on the mean) in units of $10^{-6}$. We report the values obtained for different choices of Galactic masks.}
\label{tab:mumonopole}
\end{table}
\begin{table}[t]
\begin{tabular}{cccccc}
\toprule\toprule
 $|\langle\mu\rangle| < 95\%$ C.L. $[\times 10^{-5}]$        & PL90           & PL80& PL60   & PL40& $\sigma_{syst}$  \\ 
 \midrule
$A_d+\beta_d$ &$<47$ & $<51$  & $< 47$  & $< 53$& 2.5\\
$A_d$, fixed $\beta_d$                         &$<74$ & $<50$  & $<105$ & $<179$& 48\\
$A_d+A_s$                 &$<{\bf 252}$ &$<200$&$<162$ & $<102$& 54\\
$A_d+A_{\rm ff}$               &$<{\bf 174}$ &$<120$&$<96$  & $<118$& 47\\
FIRAS residual           &$<72$   &$<57$  &$<89$  & $<147$&34\\
\bottomrule\bottomrule
\end{tabular}
\caption{$\mu$ distortion monopole upper limits for different choices of foreground separation methods and Galactic masks. We outline in bold values for which the monopole is detected at more than $2\sigma$ significance. Since it only happens Galactic masks that include a good fraction of the Galactic plane, this hints for a clear foreground contamination. The last column reports the standard deviation of the values of the upper limits obtained for different masks which we considered as an estimate of systematic uncertainty on the constraint. }
\label{tab:mumonopole-vs-fg}
\end{table}

\begin{figure}
\includegraphics[width=\columnwidth]{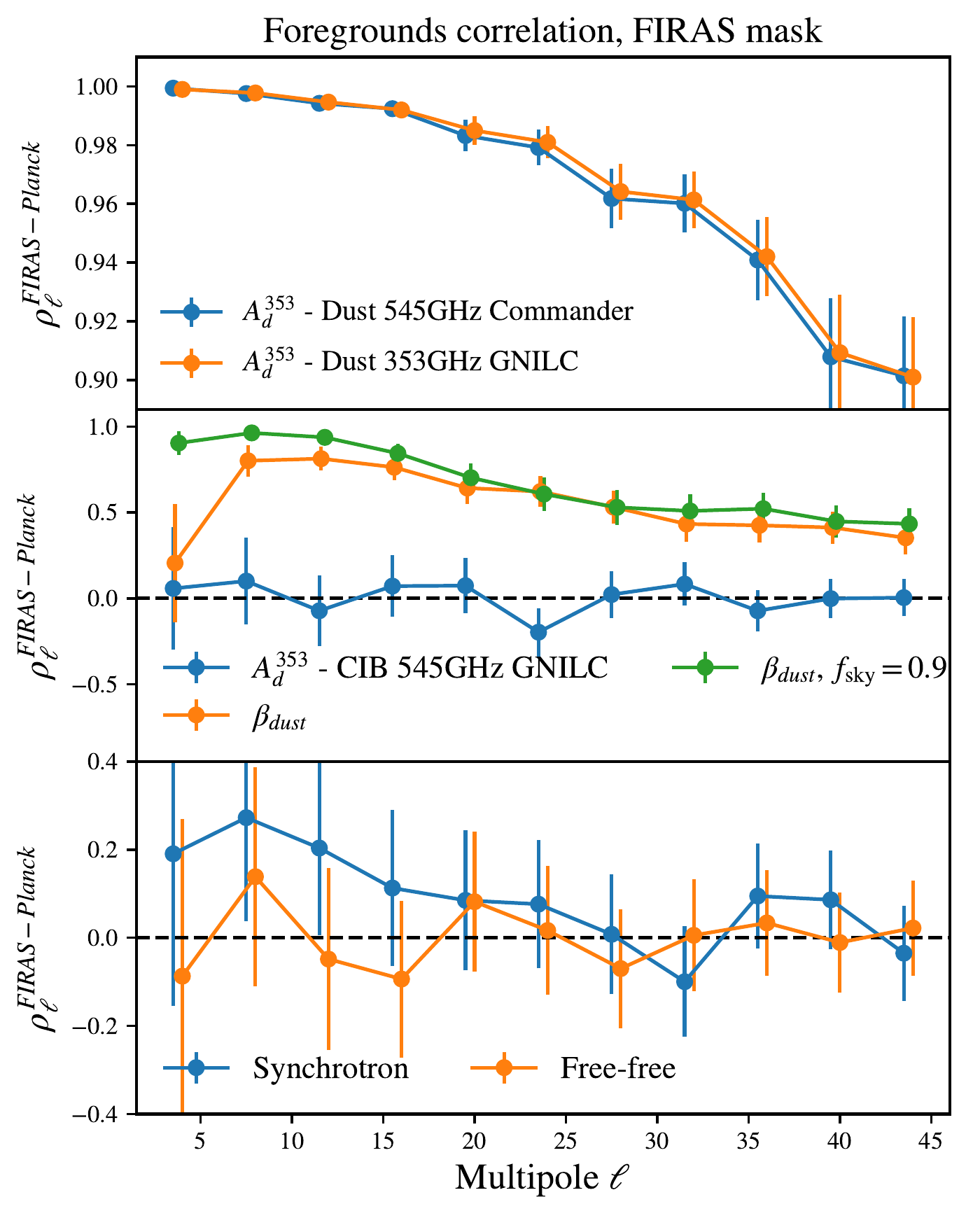}
\caption{Cross-correlation coefficient between foreground maps extracted by our baseline component separation analysis and different \planck foreground templates. Our foreground maps for the dust intensity and spectral index are highly correlated with the \planck data products across all angular scales considered in this work. Discrepancies at the largest scales in $\beta_d$ are alleviated when excluding the brightest regions of the Galactic plane (removed with a P90 mask). The low frequency foreground components (shown in the bottom panel) are conversely mainly unconstrained and their cross-correlation coefficient is consistent with 0. Points of different templates have been shifted in $\ell$ for visualization purposes.}
\label{fig:fgcorr}
\end{figure}

The $\Delta T$ template provides information on anisotropic variations of the CMB temperature. 
The dominating anisotropy is the dipole induced by the motion of the Solar system and the FIRAS satellite with respect to the CMB reference frame.  
We fit a dipolar emission to the map using the \texttt{healpy} dedicated routines on the same fiducial mask of \fixsenspec and find an amplitude of the dipole $A_{\rm dipole}=3326\,\mu K$ along a direction in Galactic coordinates $(l,b)=(264.13^\circ,49.21^\circ)$. These values are consistent within $\sim 1.5\sigma$ with the original estimate of the FIRAS team of $(l,b)^{\rm FIRAS} =(264.14^\circ \pm 0.15^\circ,48.26^\circ \pm 0.15^\circ)$ and $A^{\rm FIRAS}_{\rm dipole}=3369 \pm 40\, \mu K$ in \fixsenspec\ except for the $b$ coordinate of the dipole. The discrepancy is due to the residual correlation between $\mu$ and $\Delta T$ parameter. We can restore a complete agreement in the direction of the dipole using more aggressive Galactic masks $(l,b)^{\rm FIRAS}$ or if we do not fit for the $\mu$ distortion on the same mask. In this case we obtain $A_{\rm dipole}=3332\,\mu K$, $(l,b)=(264.30^\circ,48.08^\circ)$.
The average of the $\Delta T$ template can be added to the pivot value of $T_0=2.725$ we used in our data model to estimate the value $\hat{T}_0$ of $T_0$ on the sky. 
By doing so, we obtain $\hat{T}_0 = 2.723$ for our fiducial component separation method. 
The value is stable with respect to the mask choice and consistent with the $\approx 5$ mK temperature reduction compared to the FIRAS original estimates in e.g. F96 $T_0 = 2.728\pm 0.004$ including statistical and systematics uncertainties as well as the estimates of F96 $T_0 = 2.717\pm 0.007$ fitting the data to the CMB dipole SED. The shift in our value is expected due to the calibration systematics correction applied to the FIRAS reprocessed data in \textsc{HEALPix} pixelization.\footnote{\url{https://lambda.gsfc.nasa.gov/product/cobe/firas_tpp_info.html}}
We also compare the monopole of our dust map with the one obtained by rescaling the Commander dust intensity from the reference 545 GHz to our pivot frequency of 353 GHz using the map of the spectral index and $T_d$ fitted for the low resolution Commander maps and the emission law of Eq.~\eqref{eq:dust-emission}. 
We find a consistent value of 0.29 MJy/sr for both those maps if the \planck P90 mask is applied before averaging the pixels. 
Since our pivot frequency is different than the one chosen by Commander, a direct comparison of the mean spectral index is not straightforward. 
We obtain a mean dust spectral index across the map $\langle \beta_d\rangle = 1.22\pm 0.02$. 
To further validate our component separated products, we compute the cross-correlation coefficient between them and foreground templates released by \planck. 
Specifically, we make use of the dust $A^{545}_d$ and $\beta_d$ templates derived using the Commander and GNILC algorithms for the 2015 release of \planck data. 
We also correlate the Commander templates for free-free and synchrotron emission against our $A_s$ and $A_{\rm ff}$ maps.  
For the sake of simplicity, we use a pseudo-$C_\ell$ power spectrum estimator as implemented in the public code \texttt{NaMASTER}\footnote{\url{https://github.com/LSSTDESC/NaMaster}} \cite{master,namaster} to compute the spectra required to evaluate the cross-correlation coefficient between two $X$ and $Y$ fields, $\rho_\ell   = C_\ell^{XY}/\sqrt{C_\ell^{XX}C_\ell^{YY}}$. 
The results for our baseline $A_d+\beta_d$ foreground model are shown in Fig.~\ref{fig:fgcorr}, where we use a FDS mask and adopt the analytical Gaussian error bars on the cross-correlation coefficient \cite{louis2019}. 
Our dust amplitude products show a remarkable degree of correlation ($\gtrsim 90\%$) with the \planck dust products. 
The differences observed for $\beta_d$ are mainly confined to the largest multipoles and are driven by differences in separation in the Galactic plane. 
The cross-correlation coefficient increases to $\sim 90\%$ at large scales if we exclude the Galactic plane by applying a P90 mask in addition to the FDS mask when computing the cross-correlation coefficient.  
The CIB does not seem to contaminate our dust templates. 
The low frequency foregrounds such as free-free and synchrotron are conversely not well constrained by our component separation since the FIRAS data lack a low frequency lever arm to effectively anchor those emissions, with 60 GHz being the lowest available frequency. 
The cross-correlation coefficient is therefore consistent with the null line for both of these components. This validates our baseline choice to not include those in the reference cleaning method. 

\subsection{\label{sec:spectra}Power spectra}
\begin{figure*}[tbp]
    \centering
    \includegraphics[width=\textwidth]{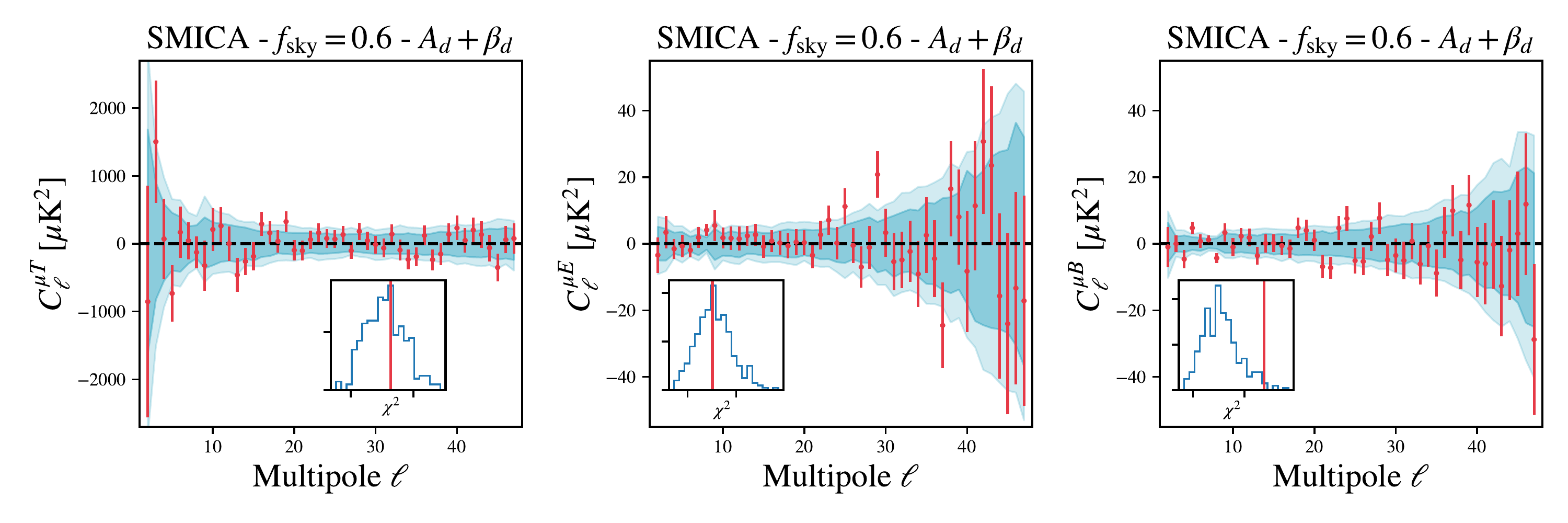}
    \includegraphics[width=\textwidth]{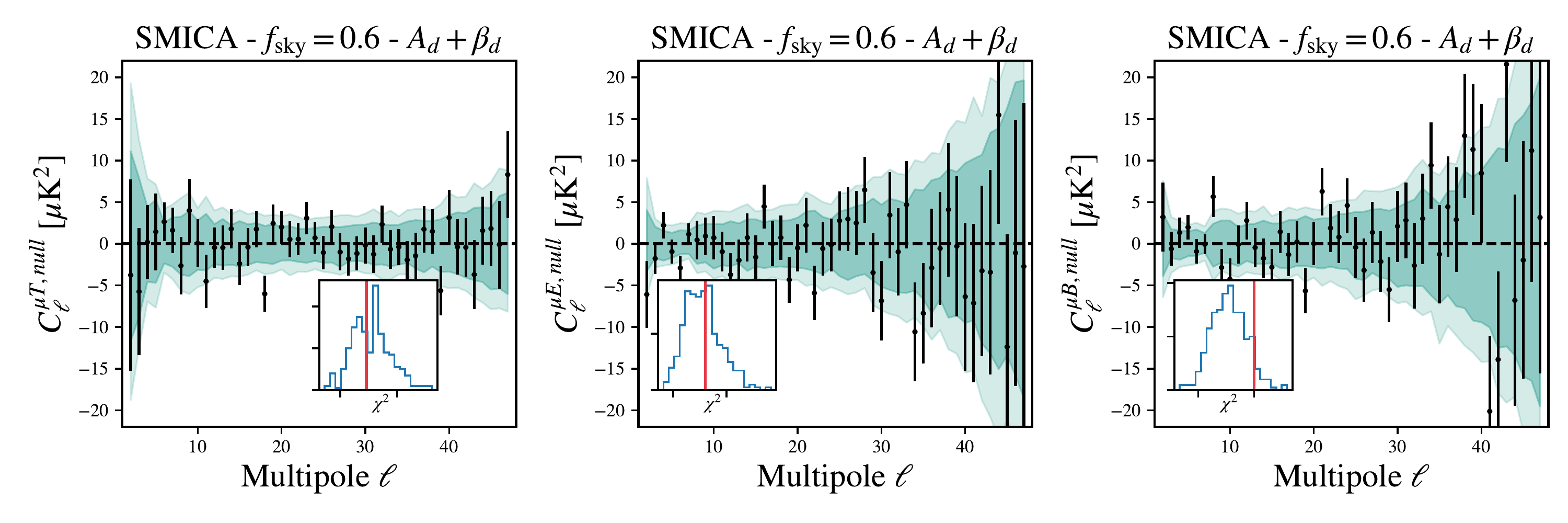}    
    \caption{\textit{Top}: cross-power spectrum between the anisotropic $\mu$-distortion map from FIRAS and the component-separated CMB temperature (left panel) and $E$-mode (right panel) maps from \textit{Planck}. This plot shows the baseline results based on the SMICA \textit{Planck}{} maps, a Galactic mask with $f_{\rm sky}=0.6$, and the Galactic dust with free amplitude and spectral index as foreground model. In each panel, the red points indicate the data while the darker and lighter shaded regions represent the 1 and 2$\sigma$ scatter in simulations that do not include any anisotropic $\mu$. The insets compare instead the $\chi^2$ statistic from simulations (blue histogram) to the value found in data (red vertical line), showing that the extracted spectra are consistent with the null line. \textit{Bottom}: same as above for the cross-correlation between $\mu$ map and the half-mission jackknife map. Data points are shown in black.}
    \label{fig:cl_mu_cross_cmb}
\end{figure*}

In Fig.~\ref{fig:cl_mu_cross_cmb} we show the second main result of this analysis, the extracted angular cross-power spectra between the FIRAS $\mu$ anisotropy map from FIRAS and the SMICA component-separated CMB temperature and $E/B$-mode polarization maps from \planck (red points).
These spectra have been extracted for our baseline foreground model (Galactic thermal dust with free amplitude and spectral index) over about 60\% of the sky using a QML approach, see Sec.~\ref{sec:ps_ext} for a detailed discussion of the method.
The dark and light blue shaded regions in Fig.~\ref{fig:cl_mu_cross_cmb} show the $(14^{\rm th},86^{\rm th})$ and $(5^{\rm th},95^{\rm th})$ percentiles of the distributions of the SD-CMB cross-spectra measured in the FFP10 simulations, respectively.
The $\mu T$ and $\mu E$ cross-power spectra recovered from FIRAS and \planck data are both consistent with zero. 
Under the null hypothesis that the $\mu$ and CMB maps are uncorrelated, we can evaluate $\chi^2_{\rm null}=\sum_{\ell\ell'}C^{\mu X}_{\ell}\Sigma_{\ell\ell'}^{-1}C^{\mu X}_{\ell'}$ and calculate the probability-to-exceed (PTE) by counting the number of simulations that have a $\chi^2_{\rm null}$ larger than that of the data.
We find PTE values of 30\% and 67\% for $\mu T$ and $\mu E$ respectively, therefore we cannot rule out the no correlation hypothesis.
The distribution of $\chi^2_{\rm null}$ for both cases are shown in the inset plots of Fig.~\ref{fig:cl_mu_cross_cmb}.

Correlation between $\mu$ distortions and B-mode of CMB polarization are potentially a unique probe of tensor non-Gaussianities on scales inaccessible by CMB bispectra. However, any tensor or mixed scalar-tensor bispectra of primordial perturbations that are isotropic are expected to leave either vanishing or strongly suppressed signatures in the observed diagonal $\langle \mu_{\ell_1} B_{\ell_1}\rangle$ correlation that are constrained by $\clmub$ \citep{orlando21}. Nonetheless we report the amplitude of $\clmub$  from FIRAS and \planck data in Fig.~\ref{fig:cl_mu_cross_cmb} 
A similar calculation to the one done for $\clmut$ and $\clmue$ yields a PTE of about 6\%, consistent with a non detection assumption. Such null test can be used to constrain parity violating mechanisms and statistical anisotropies. However, such constraints can be set more naturally in terms of amplitude of the bipolar spherical harmonics coefficients $C_{\ell_1\ell_2}^{\mu B}$ \cite{biposh} that capture off-diagonal elements of the $\mu$B correlation. A detailed analysis of such correlations together as their implications for non-Gaussian primordial scalar and tensor bispectra is left for future work. In general we note that the use of independent datasets such as FIRAS and \planck to constrain $\mu$B and $\mu$E correlations has the advantage to reduce the impact of large-scale systematics that affect the \planck data \cite{sroll2,planck20_npipe}.

To further validate the analysis, in the three lower panels in Fig.~\ref{fig:cl_mu_cross_cmb} we show the corresponding null spectra constructed by correlating the FIRAS $\mu$ map with the half-mission $T,E,B$ jackknife maps from \planck.
In this case too, the spectra are statistically consistent with the null line, yielding PTEs of 56\%,42\%, and 7\% for $\mu T$, $\mu E$, and $\mu B$ respectively. 
This test also validates the noise model since the error bars of the null spectra do not contain any CMB cosmic variance.

\subsection{\label{sec:systematics}Robustness tests}
In this section we perform different systematic tests to validate and assess the robustness of our cross-correlation measurements.

\subsubsection{Stability against foreground models}
\begin{figure*}[!htbp]
\includegraphics[width=\textwidth]{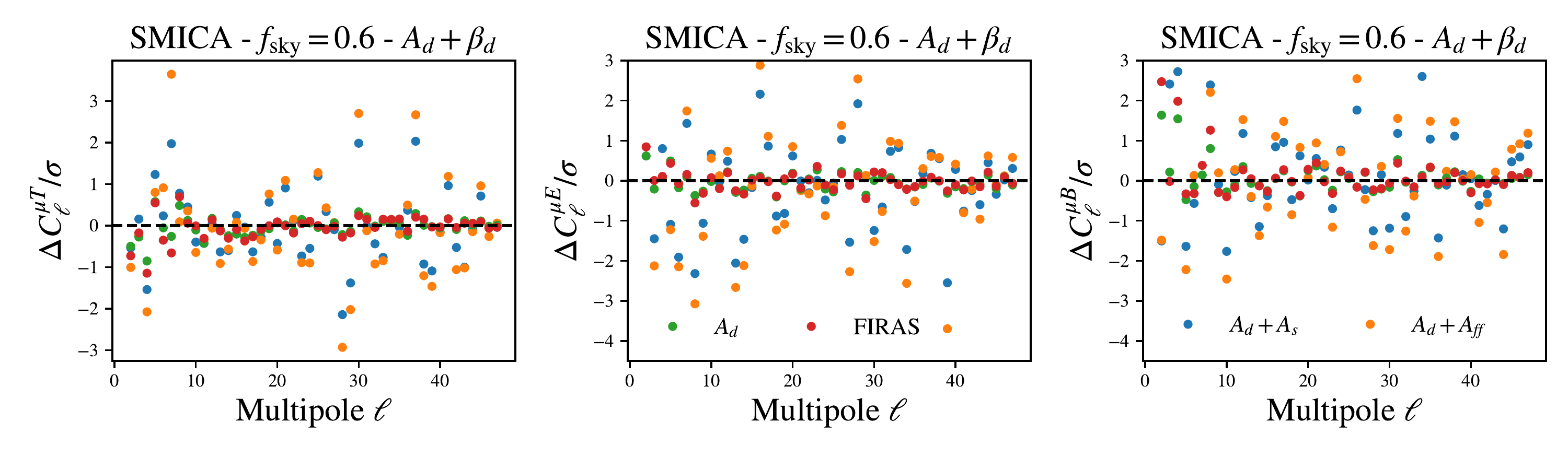}\\
\includegraphics[width=\textwidth]{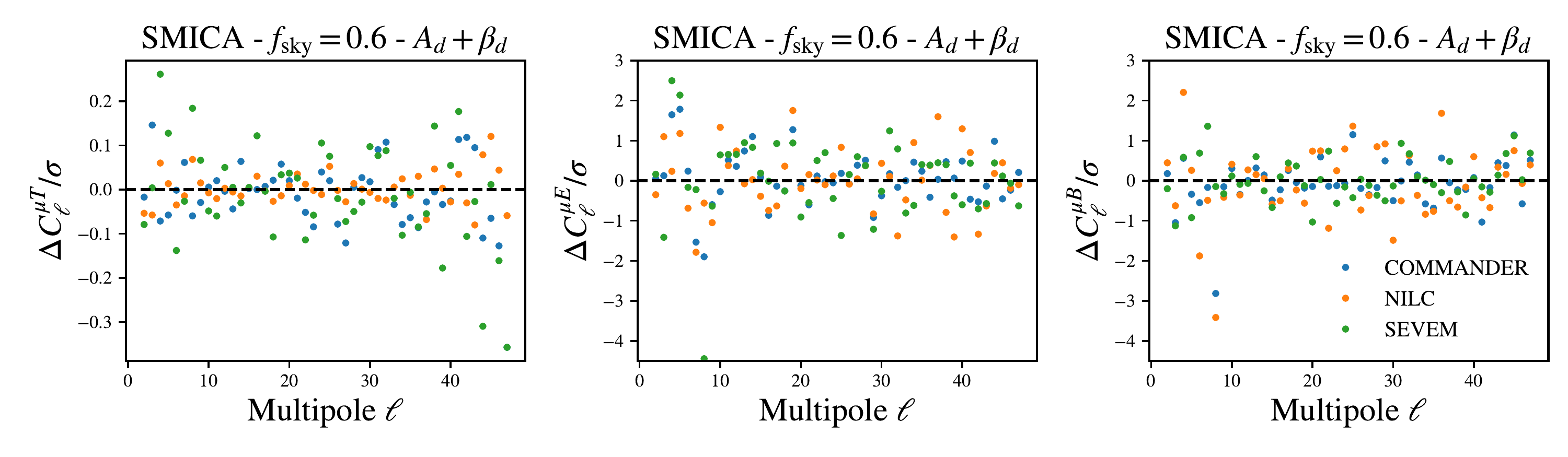}
\caption{Top: changes in the cross-correlation power spectrum between $\mu$ distortions and CMB anisotropies of the \planck SMICA maps for different component separation methods relative to the statistical uncertainty of our baseline analysis. For the majority of the multipoles the fluctuations are within $1\sigma$. Bottom: changes in the cross-correlation power spectra used in this work when the reference SMICA maps are swapped with different \planck CMB maps obtained adopting different component separation algorithms. Fluctuations in $\clmut$ are marginal but some multipoles display fluctuations above $3\sigma$ compared to our baseline case for $\clmue,\clmub$. However, these are consistent with noise-induced fluctuations.} 
\label{fig:stability}
\end{figure*}

We check the stability of our cross-power spectra measurements with respect to different component separation methods both in the CMB and the $\mu$-distortion side. 
We show the summary of our results in Fig.~\ref{fig:stability}. 
On the CMB side, we compare the changes in the $\clmut, \clmue, \clmub$ power spectra obtained using different component separated maps provided by the \planck collaboration (see  Sec.~\ref{sec:planckdata}) relative to the statistical uncertainty determined in our baseline analysis with the SMICA map. 
Shifts in the $\clmut$ values are minor and well below $0.3\sigma$. 
The cross-power spectra involving polarization show larger variations that remain below $2\sigma$ for most of the multipoles.  
We check that these shifts are expected given the noise level of the \planck maps. For this purpose, 
we generate a set of MC simulations where, for a specific component separation method, we add a CMB realization (common to all the component separation methods) to each noise realization of the FFP10 suite for that specific method. 
We then cross-correlate them with a realization of the $\mu$ map generated from its pixel-pixel covariance, both using the full and diagonal approximations. 
This allows us to have a consistent estimate of the power spectrum using the hybrid approach used to analyze the real data. 
From this set of simulations, we compute the covariance of the differences of the bandpowers between CMB component separation methods that we then use to compute the $\chi^2$ statistics for each of the MC realizations. We then compare the $\chi^2$ value obtained from the shifts in bandpowers measured on data with the $\chi^2$ distribution obtained from simulations and find that the observed fluctuations between data using different CMB component separation methods are consistent with shifts induced by the noise (PTE $>5$\%).

We repeat a similar analysis to test whether the SD-CMB spectra obtained using different foreground cleaning assumptions in the construction of the $\mu$ map lead to statistically consistent results when correlated with the SMICA map. 
For this purpose we generate a MC simulation set sharing the same CMB for each foreground cleaning model of Sec.~\ref{sec:fgmodels} and independent realizations of the $\mu$ map, since we consider the noise to be mostly uncorrelated between different component separation methods as discussed in Sec.~\ref{sec:mucov}. 
We find that for the $A_d+A_{\rm ff}$ model, which displays deviations up to $3\sigma$ for $\clmue$ and $\clmub$ when compared to our baseline analysis, the PTE for the spectra are consistent at the level of 75\% and 73\% respectively. This test shows that all the methods clean with good consistency the major foreground at this frequency and angular scales, i.e. the Galactic dust, and that residuals of the cleaning do not matter much for the cross-correlation. Different methods however do show differences in the final constraints on $\fnl$ and in the overall monopole since the residual noise and signal are different. The baseline model $A_d+\beta_d$ delivers the most stable results against changes in the sky mask or CMB map used in the analysis as shown in Tab.~\ref{tab:mumonopole-vs-fg} and \ref{tab:fNL_constraints}, which is why we use this as our baseline analysis. 

\subsubsection{Analysis choices}
We check the stability of our analysis choice with respect to several assumptions made in the analysis and in the construction of our reference data model of Eq.~\eqref{eq:data-model}. In particular we change the value $T=T_0$ where we evaluate the gradient of the blackbody spectrum to the original value used by FIRAS $T_0=2.728$; in addition we also try to simultaneously fit for the $y$ and $\mu$ distortion. Changing the value of $T_0$ mainly influences the value of the fitted CMB anisotropy map $\Delta T$ but does not lead to any appreciable change in the value of the cross-correlation power spectra nor in the monopole. Averaging the new $\Delta T$ maps and summing it to the new value gives consistent results with the estimate we provide in Sec.~\ref{sec:results}.

Another plausible concern is the fact that we neglect the possible interplay of $y$ and $\mu$ distortions as we fit for $\mu$ only. A cross-correlation between $\mu$ and $y$ distortions could itself be used to constrain non-Gaussianities on different scales and configurations \cite{chluba16a}.
We do not expect $y$-type distortions, whose leading term comes from astrophysical objects, to heavily contaminate the measurements at large angular scales in the cross-correlation power spectra. Its anisotropies are in fact mainly sourced by the emission of single massive objects and thus thier power spectrum grows at higher multipoles \cite{Komatsu:2002wc}. 
As a sanity check, we rerun our analysis pipeline when both $\mu$ and $y$ distortions are allowed to be present. 
We find the two parameters to be significantly anticorrelated (by more than 80\% on average), suggesting the lack of constraining power of the dataset on both the distortion types simultaneously. 
We correlate the resulting $\mu$ map with the $T$ anisotropy of \planck, which is the one potentially contaminated by the tSZ emission, and find no shift in the $\clmut$ bandpowers compared to our baseline setup.
Finally, we check the stability of our results with respect to the analysis mask. Our monopole constraint are robust to the choice of the mask. We find the $\clmut$ and $\clmue$ power spectra are still consistent with zero with a reduced statistical uncertainty, however our final cosmological constraint are consistent with the result we present in Sec.~\ref{sec:fNL_constaints} due to a slight shift in the central value that might hint for the presence of mismodeled noise or minor foreground residuals.

\subsubsection{Foreground deprojection}\label{sec:mu-fgdeproj}
We test for the presence of residual foreground emission in the $\mu$ map by deprojecting external templates of Galactic and extragalactic foregrounds. 
If we assume that the $\mu$ map is contaminated by different foreground emissions described by a set of templates $f^{i}$ of amplitude $\alpha_{fg}^{i}$, we can write 
\begin{equation}
\mu = \mu + \sum_{i} \alpha_{fg}^{i}f^{i}.
\end{equation}
Assuming the templates are uncorrelated, we can estimate the amplitude of the single template as $\alpha_{fg}^{i} = C_\ell^{\mu f^{(i)}}/C_\ell^{f^{i}f^{i}}$ and derive the foreground-free power spectrum of the $\mu$ map from the measured power spectrum  $\hat{C}_\ell^{\mu\mu}$ as
\begin{equation}
C_\ell^{\mu\mu} =\hat{C}_\ell^{\mu\mu} \left(1-\sum_i\left(\rho_\ell^{\mu f^{i}}\right)^2\right),
\end{equation}
where $\rho_\ell^{\mu f^{(i)}}$ is the cross-correlation coefficient between the foreground template and the $\mu$ map and $\hat{C}_\ell^{f^{(i)}f^{(i)}}$ the measured power spectrum of the template. We can extend the formalism to cross-spectra with a CMB map $X$ where we assume we have contaminations described by the same template with a different amplitude $\beta_{fg}^{i}$ such that
\begin{equation}
X = X + \sum_{i} \beta_{fg}^{i}f^{i} \qquad X\in [T,Q,U],
\end{equation}
so that the foreground deprojected cross-correlation power spectrum reads
\begin{equation}
C_\ell^{\mu X} =\hat{C}_\ell^{\mu X} - \sum_i \frac{ \hat{C}_\ell^{\mu f^{i}}\hat{C}_\ell^{X f^{i}}}{\hat{C}_\ell^{f^{i}f^{i}}} \qquad X\in [T,E,B].
\end{equation}
We perform this template deprojection for different types of foreground using \planck-based templates that were degraded to the FIRAS resolution and smoothed with the FIRAS effective beam. 
As for the data pipeline, we compute the cross-spectra using the \xqml power spectrum estimation pipeline. 
For the Galactic foregrounds we use the Commander 2015 dust, synchrotron, free-free and CO(2-1) line templates while for the extragalactic emissions we use both Compton-$y$ maps provided by the \planck collaboration, that are based on different component separation methods, and the GNILC CIB emission templates at 353 GHz. In all cases we do not find significant shifts in the bandpowers, which fluctuate well within the statistical error bars after foreground deprojection. 

\section{\label{sec:fNL_constaints} $\fnl$ constraints}
In this section we translate the extracted SD-CMB cross-power spectra presented in Sec.~\ref{sec:results} into constraints on the amplitude of the local-type primordial non-Gaussianities at small scales.

Let us first review the theoretical modelling of the cross-correlation signal between $\mu$ fluctuations and primary CMB temperature and polarization anisotropies.  
To highlight the connection of the SD-CMB cross-spectra to primordial non-Gaussianities, we closely follow the earlier works of \cite{pajer12,ganc12,ota16,ravenni17} and define the so-called local-form bispectrum as
\begin{equation}\label{eq:bispectrum}
    B(k_1,k_2,k_3) = -\frac{6}{5} \fnl^{\rm loc} \left[ P_{\zeta}(k_1)P_{\zeta}(k_2) + 2 \, \text{perm.}\right],
\end{equation}
where $P_{\zeta}(k)$ is the primordial spectrum of curvature perturbation, $P_{\zeta}(k)\propto k^{n_s-4}$.
This bispectrum peaks in the so-called squeezed triangle configuration ($k_L\equiv k_1\ll k_2 \approx k_3 \equiv k_s$)\footnote{For which we obtain $B(k_1,k_2,k_3)\to B(k_s,k_s,k_L)\approx -\frac{12}{5}\fnl P_{\zeta}(k_s)P_{\zeta}(k_L)$.} and is interesting to study for a twofold reason. 
First, because measuring a statistically significant deviation of $\fnl$ from zero would disfavour single-field inflation models \citep[e.g.,][]{maldacena03,acquaviva03} and second, because this shape of bispectrum is less prone to contaminations from late-time effects, such as the lensing-integrated Sachs-Wolfe effect bispectrum, which effectively makes it a robust probe of the early universe \citep[e.g.,][]{goldberg99}.

The spherical harmonic coefficients of the primary CMB and $\mu$-type distortion fluctuations are linked to the primordial curvature perturbation $\zeta(\bf k)$ through
\begin{align}
a_{\ell m}^{X}&=4 \pi i^{\ell} \int \frac{\mathrm{d}^{3} \boldsymbol{k}}{(2 \pi)^{3}} \mathcal{T}_{\ell}^{X}(k) Y_{\ell}^{m *}(\hat{\boldsymbol{k}}) \zeta(\boldsymbol{k})\\
a_{\ell m}^{\mu}&=
    \begin{aligned}[t]
    &4 \pi(-i)^{\ell} \int \frac{\mathrm{d}^{3} \boldsymbol{k}_{\mathbf{1}}}{(2 \pi)^{3}} \frac{\mathrm{d}^{3} \boldsymbol{k}_{\mathbf{2}}}{(2 \pi)^{3}} \mathrm{~d}^{3} \boldsymbol{k}_{\mathbf{3}} \delta^{(3)}\left(\boldsymbol{k}_{\mathbf{1}}+\boldsymbol{k}_{\mathbf{2}}+\boldsymbol{k}_{\mathbf{3}}\right) \\
& \times Y_{\ell}^{m *}\left(\hat{\boldsymbol{k}}_{\mathbf{3}}\right) j_{\ell}\left(k_{3} r_{\mathrm{ls}}\right) f^{\mu}\left(k_{1}, k_{2}, k_{3}\right) \zeta\left(\boldsymbol{k}_{\mathbf{1}}\right) \zeta\left(\boldsymbol{k}_{\mathbf{2}}\right),
    \end{aligned}
\end{align}
where $X=\{T,E\}$ denotes either temperature or $E$-mode polarization, $\mathcal{T}_{\ell}^{X}$ and $f^{\mu}$ are the radiation and $\mu$-mode transfer functions respectively, $j_{\ell}(x)$ is the spherical Bessel function, and $r_{\rm ls}\approx$ 14 Gpc is the comoving distance to the last-scattering surface. 
We calculate the radiation transfer function using CAMB \citep{lewis00} and remind the reader that the $\mu$ window function picks up signal from the  primordial scalar power spectrum in the $50 \lesssim k \lesssim 12000$ Mpc$^{-1}$ range \citep[following e.g.,][]{chluba16a}.

Then, it is straightforward to calculate the SD-CMB cross-spectrum (in the squeezed limit) as 
\begin{equation}\label{eq:SD-CMB-ps}
\begin{aligned}
    C_{\ell}^{\mu X} &\approx-4 \pi \frac{12}{5} \int \frac{k^{2} \mathrm{~d} k}{2 \pi^{2}} \mathcal{T}_{\ell}^{X}(k) j_{\ell^{\prime}}\left(k r_{\mathrm{ls}}\right) P_{\zeta}(k) \\
    &\times\int \frac{q_{1}^{2} \mathrm{~d} q_{1}}{2 \pi^{2}} f^{\mu}\left(q_{1}, q_{1}, k\right) P_{\zeta}\left(q_{1}\right).
\end{aligned}
\end{equation}
Note that the second integral in the equation above is approximately equivalent to the definition of the monopole of the $\mu$ distortion, which we set to its $\Lambda$CDM expectation, $\langle\mu\rangle= 2.3\times 10^{-8}$ \cite{chluba16b}.
From Eq.~\eqref{eq:SD-CMB-ps}, we can see that the SD-CMB cross-power spectra linearly depend on the product of the $\fnl$ parameter and $\langle\mu\rangle$, i.e. $\langle a_{\ell m}^{\mu} a_{\ell m}^{X}\rangle \propto \fnl\langle\mu\rangle$.
Therefore, a larger value of $\langle\mu\rangle$ would translate to a tighter constraint on $\fnl$. 

\begin{figure}[!tbp]
    \centering
    \includegraphics[width=\columnwidth]{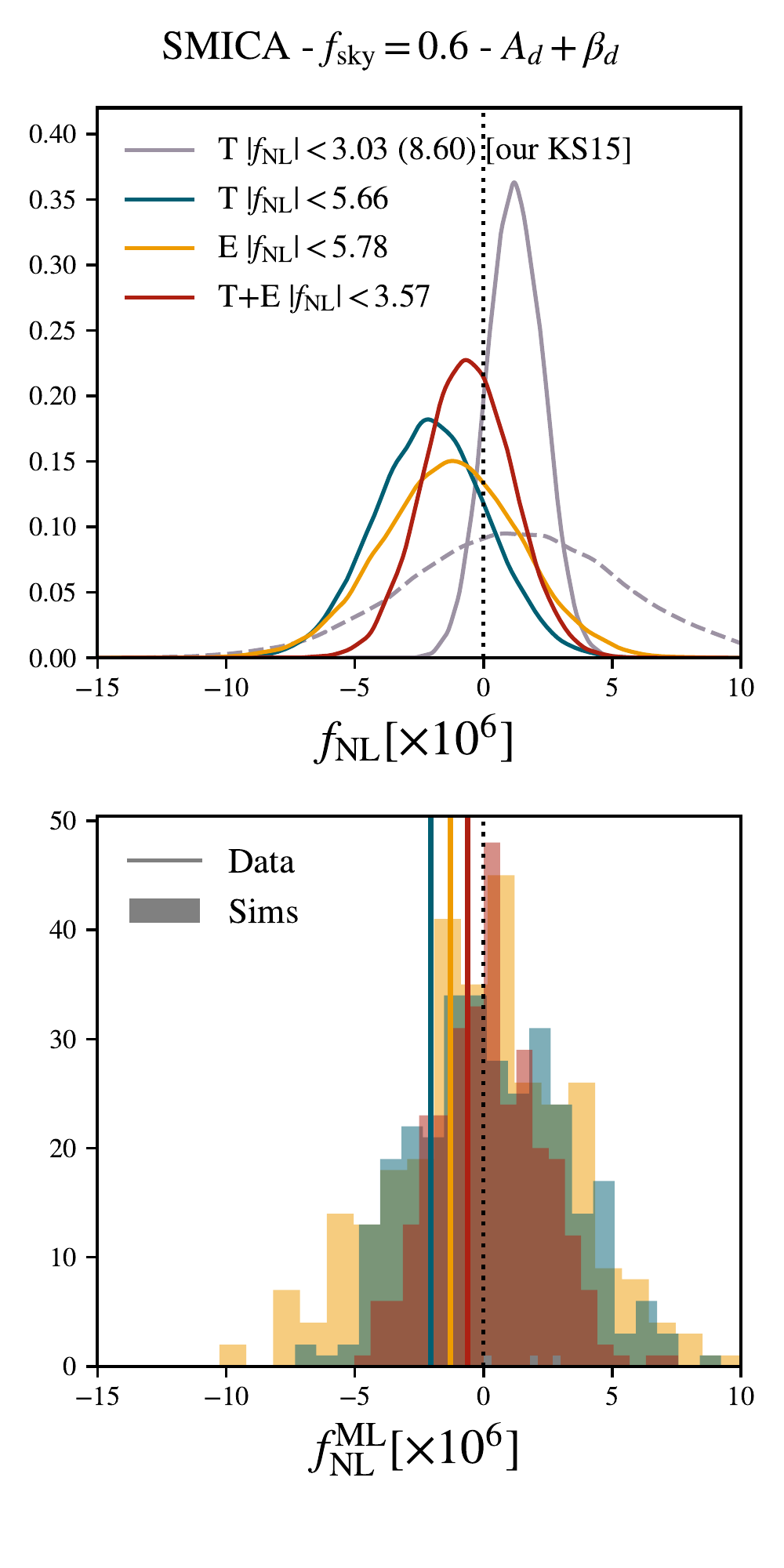}
    \caption{\textit{Top}: Posteriors on $\fnl$ from $\mu T$ (blue), $\mu E$ (yellow), and their joint analysis (red). The grey lines show the constraints on $\fnl$ from our reprocessing of the \citet{khatri15} data. In particular, the solid one only reflects the effects of the proper band power uncertainties estimation, while the dashed line also accounts for the statistical and systematic uncertainties related to the CIB offset residuals. The numbers in the legend show the corresponding $2\sigma$ upper limits on the absolute value of $\fnl$ (the number in parenthesis corresponds to the pessimistic constraint from the KS15 reanalysis). \textit{Bottom}: Maximum-likelihood $\fnl$ values recovered from the data (vertical lines) and from \planck FFP10 simulations (histograms). The color coding is the same as the top panel.}
    \label{fig:fNL_constraints}
\end{figure}

We now have all the tools needed to carry out the cosmological inference.
First, we set the amplitude of the primordial scalar perturbation power spectrum to $A_s = 2\times 10^{-9}$ and its spectral index to $n_s=0.965$, and calculate the $\mu T$ and $\mu E$ template spectra using the theoretical framework outlined above.
We then compare the SD-CMB cross-spectra measured in our baseline setup over the whole $\ell$-range ($2\le \ell \le 47$) to the theoretical curves following the analysis scheme discussed in Sec.~\ref{sec:fNL_fit}.
The results are presented in Fig.~\ref{fig:fNL_constraints}.

In the top panel, we show the one-dimensional $\fnl$ posteriors obtained by sampling the Gaussian likelihood in Eq.~\eqref{eq:spectra_like} assuming an unbound uniform prior on $\fnl$.  
The temperature-only analysis (blue curve) yields a 95\% upper limit on the absolute value of $|\fnl|$ of \fnlT, the $E$-mode polarization spectra (yellow curve) reveals $|\fnl| < \fnlE$, and their joint analysis suggests $|\fnl| < \fnlTE$.
The constraints on $\fnl$ from SD-CMB cross-spectra are robust against changes in the component-separation algorithm used to clean the \planck CMB temperature and polarization maps. 
In Tab.~\ref{tab:fNL_constraints}, we summarize the results based on the alternative cleaning methods which are at the level of few $\times 10^6$, similar to those from SMICA algorithm.
If we instead want to be agnostic about the value of the $\mu$ monopole, and thus set model-independent constraints, we can directly infer upper limits on the $\fnl\langle\mu\rangle$ product. 
The corresponding $2\sigma$ constraints from $\mu T$, $\mu E$, and their joint analysis yield $\fnl\langle\mu\rangle < 0.13$, $<0.15$, and $<0.08$ respectively. We remind that for high values of $\fnl$, the expressions we used for $\clmut$, $\clmue$ become inaccurate as they are based on perturbative expansions in primordial fluctuations. Improving the theoretical modelling of these signals  is beyond the scope of thus work but we justify their use given the high-noise regime of our measurements and the consistency with 0 of all the measured spectra.
To complement the analysis, in the lower panel of Fig.~\ref{fig:fNL_constraints} we compare the maximum-likelihood values of $\fnl^{\rm ML}$ found for the data (vertical lines) to those obtained from the FFP10 simulation cross-spectra (histograms).
Two things are worth noting. 
First, the maximum-likelihood values are in agreement with the central values from the Bayesian analysis reported in the top panel.
Second, for each of the $\mu T$, $\mu E$, and $\mu T + \mu E$ cases separately, the best-fit value from the data is consistent with the distribution of $\fnl$ from the simulations that do not contain any primordial local-type non-Gaussian signal.

\begin{figure*}[!htbp]
    \centering
    \includegraphics[width=\textwidth]{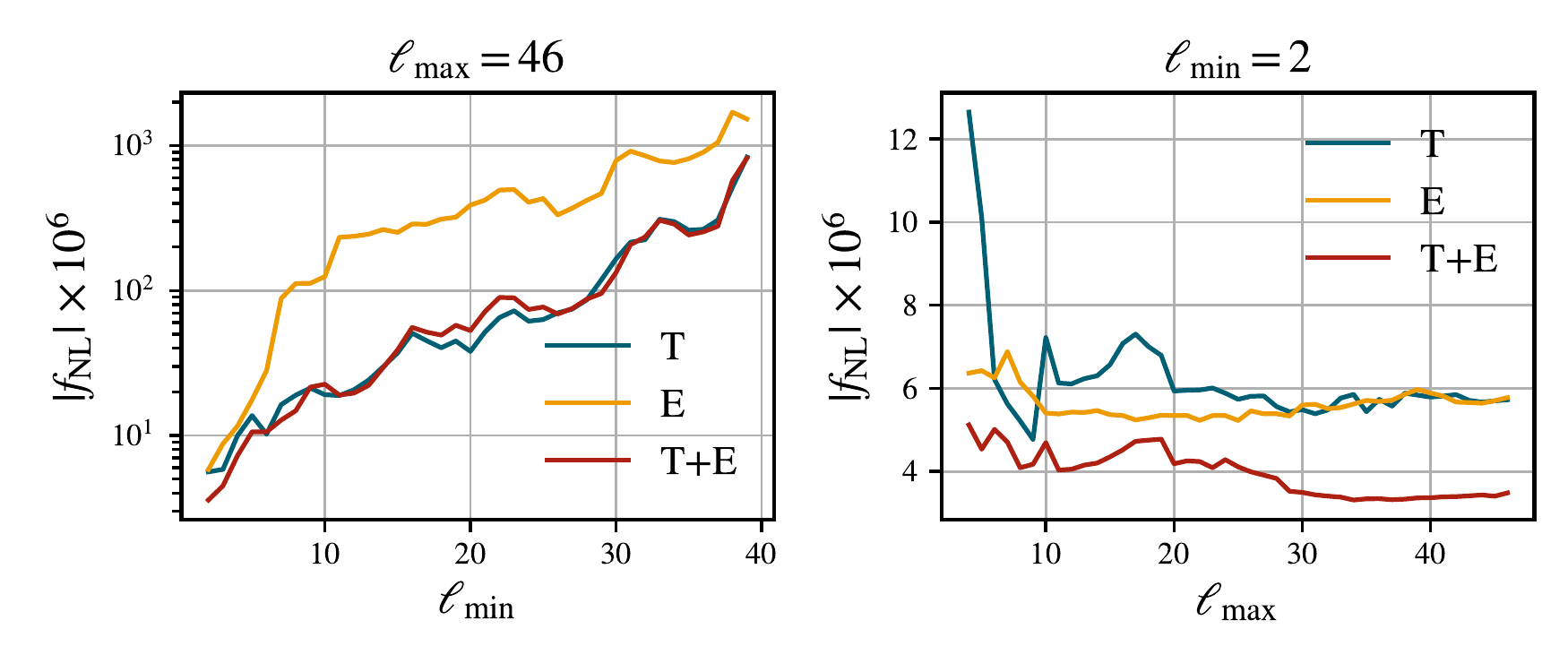}
    \caption{$2\sigma$ upper limits on $|\fnl|$ as function of the minimum multipole $\ell_{\rm min}$ when fixing $\ell_{\rm max}=46$ (left panel) and as function of the maximum multipole when setting $\ell_{\rm min}=2$. These curves are reported for our baseline analysis. In each panel, the blue, yellow, and red lines show the constraints from $\mu T$, $\mu E$, and their joint analysis respectively.}
    \label{fig:fNL_constraints_lmin_lmax}
\end{figure*}

Finally, in Fig.~\ref{fig:fNL_constraints_lmin_lmax} we show the $2\sigma$ upper limits on $|\fnl|$ for our baseline case as function of the minimum and maximum multipole included in the likelihood analysis (left and right panel respectively).
As we can see, most of the constraining power comes from the first $\ell \lesssim 20$ multipoles, as the upper limit curves flatten out for larger multipoles.
For example, reducing the minimum multipole from $\ell_{\rm min}=2$ to 10 degrades the $2\sigma$ upper limit on $|\fnl|$ by about a factor 6 for both $\mu T$ and $\mu E$, and roughly 3.6 for their combined analysis.

The upper limits on $\fnl$ found in this section are significantly weaker than those from the analysis of the \planck bispectra, $|\fnl|\lesssim 5$ \citep{planck18_ng}, or those from galaxy surveys, $|\fnl|\lesssim 20$ \citep[e.g.,][]{mueller21}.
However, we emphasize that SD-CMB correlations are sensitive to $\fnl$ at wavenumbers $k_{\mu}\approx 740$ Mpc$^{-1}$ which are much smaller than those probed by CMB temperature and polarization bispectra or large-scale structure. 
This allows us to place constraints on the scale-dependence of non-Gaussianities \citep[e.g.,][]{sefusatti09,byrnes10}. 
Considering a simple phenomenological model of the running of primordial non-Gaussianities, $\fnl(k)\simeq \fnl(k_0)\left(k/k_0\right)^{\nnl}$ with $k_0\approx 0.05$ Mpc$^{-1}$ and $\fnl(k_0)\approx 5$, our limit on $|\fnl(k_{\mu})|\lesssim 3\times 10^{6}$ translates to $\nnl\lesssim 1.4$.
\begin{table}[]
\begin{tabular}{c|cccc}
\toprule\toprule
                & SMICA & COMMANDER & NILC & SEVEM \\
\midrule\midrule
\rowcolor{gray!20}\multicolumn{5}{c}{$A_d + \beta_d$}                \\ 
\midrule
$\mu T$         & $\mathbf{< 5.66}$ & $< 5.06$  & $< 6.53$ & $< 5.19$\\
$\mu E$         & $\mathbf{< 5.78}$ & $< 8.35$  & $< 6.95$ & $< 7.09$\\
$\mu T + \mu E$ & $\mathbf{< 3.57}$ & $< 3.61$  & $< 4.18$ & $< 3.84$\\ 
\midrule
\rowcolor{gray!20}\multicolumn{5}{c}{$A_d$}                          \\ 
\midrule
$\mu T$         & $< 5.82$      &$< 9.58$   &$< 9.57$    & $< 8.38$\\
$\mu E$         & $< 5.37$      &$< 6.85$   &$< 5.08$    & $< 5.50$\\
$\mu T + \mu E$ & $< 3.46$      &$< 5.21$   &$< 7.15$    & $< 5.26$\\ 
\midrule
\rowcolor{gray!20}\rowcolor{gray!20}\multicolumn{5}{c}{FIRAS}                          \\ 
\midrule
$\mu T$         & $< 11.49$     &$< 10.55$           &$< 7.72$      &$< 8.27$       \\
$\mu E$         &$< 7.67$       &$< 6.96$           &$< 5.63$      &$< 5.54$       \\
$\mu T + \mu E$ &$< 7.66$       &$< 5.45$           &$< 5.09$      &$< 4.53$       \\ 
\midrule
\rowcolor{gray!20}\multicolumn{5}{c}{$A_d + A_s$}                    \\ 
\midrule
$\mu T$         &$< 6.56$       &$< 6.01$ & $< 7.84$      &$< 8.61$       \\
$\mu E$         &$< 11.39$      &$< 8.98$           &$< 12.57$      &$< 9.36$       \\
$\mu T + \mu E$ &$< 5.86$       &$< 4.17$           &$< 5.61$      &$< 7.05$       \\ 
\midrule
\rowcolor{gray!20}\multicolumn{5}{c}{$A_d + A_{\rm ff}$}                    \\ 
\midrule
$\mu T$         &$< 8.19$       &$< 7.36$           &$< 9.51$      &$< 11.63$       \\
$\mu E$         &$< 9.88$       &$<10.04$           &$< 10.03$      &$< 8.71$       \\
$\mu T + \mu E$ &$< 6.53$       &$< 7.10$           &$< 7.77$      &$< 7.62$       \\ 
\bottomrule\bottomrule
\end{tabular}
\caption{95\% upper limits on $|\fnl| \times 10^6$ from the cross-power spectra between anisotropic $\mu$ and CMB temperature and $E$-mode polarization anisotropies. Results are shown for different foreground models and component separation algorithms. The baseline results are highlighted in bold.}
\label{tab:fNL_constraints}
\end{table}

\section{\label{sec:khatri} Comparison with previous results}
FIRAS is still a unique dataset to constrain the $\mu$ distortion monopole. Being effectively insensitive to the absolute signal level, modern imagers like \planck often rely on the knowledge of the CMB monopole measured by FIRAS and on the annual modulation of the CMB dipole anisotropy (Solar dipole) induced by the motion of the spacecraft (orbital dipole), which is known very well, to achieve an accurate intercalibration of the frequency channels \cite{planck15_hfi,planck15_lfi}.  
At present, the precision of this technique approaches $10^{-4}$ \cite{planck20_npipe}, not far from the precision of the $T_0$ measurement \cite{fixsen2009}. Recent works suggested that much better precision needs to be achieved to reach the detection limit of expected cosmological signal of the $\mu$ distortion \cite{Mukherjee:2019pcq}. Nonetheless, some attempts have been made to extract maps of the $\mu$ distortion anisotropies from the \planck data \cite[][hereafter \ks]{khatri15} with dedicated parametric component separation algorithm called LIL \cite{Khatri:2014sra}. The data of the \ks analysis are publicly available\footnote{\url{https://theory.tifr.res.in/~khatri/muresults/}} and have been used to constrain $\fnl\lesssim 3.3\times 10^{5}$ and $\tau_{\rm NL}<2.5\times 10^{11}$ at 95\% confidence level.\footnote{We note that these limits differ from those quoted in the abstract of \ks because we inserted the exact measured values of the $\clmut$ and $\clmumu$ reported in their equations 4.5 and 4.3 in their final formulae of equation 5.32 and 5.33.} In order to perform an accurate comparison with our results, we reanalyzed the \ks data using their fiducial power spectrum estimation pipeline retaining $\fsky=0.62$, comparable to our baseline setup. We discovered several issues with their original results, that we summarize below, and derive new $\fnl$ constraints from the same dataset.

\subsection{Power spectrum and error bar estimation}\label{sec:khatri-ps}
In order to avoid noise correlation leading to a noise bias in the cross-correlation, \ks created two sets of $\mu$ maps from the half-ring splits and used the publicly available code \textsc{PolSPICE} \cite{polspice,polspice-cov} to compute $\clmut$ and $\clmumu$ where the CMB leg is given by the official \planck CMB SMICA map of the 2015 release. Given their fiducial analysis mask and maps, we are able to recover their quoted $\mu T$ cross-correlation amplitude if we correlate the \ks $\mu$ map obtained with the second half-ring split $\mu^{(2)}$ with the \planck DR2 SMICA map from the first half-ring split $T^{\rm SMICA,(1)}$. We report our findings in Table~\ref{tab:khatri}. The half-ring maps cross-spectra method ensure the absence of any noise bias in the resulting power spectra as the datasets are fully independent. Since all the band powers at scales $\ell>26$ largely deviate from zero, the authors discarded them and used only the first bin in multipoles which includes angular scales $\ell\in [2,26]$ to get cosmological constraints. We note that a full analysis of the dataset should use the cross-spectrum computed with the half-ring split 1 for the $\mu$ map and the half-ring split 2 for the $T$ map. 
This is however slightly inconsistent with zero at about $2\sigma$ level and would degrade the upper limit of $\fnl$ by pushing the measured bandpower high. The fiducial analysis in \ks adopted the analytical error bar output of the \textsc{PolSPICE} code, which assumes uncorrelated Gaussian noise and includes the Gaussian sample variance accounting analytically for the effect of the mask. However, it is unclear whether those approximations are valid for the covariance of component-separated data, in particular for their largest angular scales where the impact of non-Gaussian correlated $1/f$ noise in the data is larger and, in general because the expected distribution of the $C_\ell$ deviates from the Gaussian approximation for the power spectrum estimation methods they employed \cite{brown2004,hml2008}.  
We therefore adopt an alternative jackknifing approach to compute the error bars on the measurement. 
We first identify $N_{\rm JK}=192$ patches, corresponding to the pixel area of an \textsc{HEALPix} map of $N_{\rm side}=4$ resolution, and then remove one of them before computing the $\clmut$ or $\clmumu$ power spectra from the remaining $N_{\rm JK}-1$ patches. 
We repeat the process until every patch has been discarded once from the measurement and then compute the covariance of the measurement from all the measured $C_\ell$-s \cite{makiya2018}. 
We find that the covariance computed for all the combination of the splits exceed significantly the analytical error bars, by as much as $30\%$  for first power spectrum bin in $\clmut$ and on average by a factor $50\%$ across the whole range of multiples considered in \ks. 
Conversely, the error bars for the $\clmumu$ power spectrum, computed in \ks as the cross-correlation of the $\mu^{(1)}$ and $\mu^{(2)}$ maps is consistent with the analytical estimate in the first multipole bin but should be inflated by roughly by factor of 5 on average for the smallest angular scales. 
\begin{table}[t]
\begin{tabular}{ccccccc}
\toprule\toprule
%$[\times 10^{-12}K]$        & SMICA DR2       & SMICA DR3-- noSZ& NILC& COMM.   & SEVEM  \\ 
$[\times 10^{-12}K]$        & SMICA DR2       & SMICA DR3noSZ \\ 
 \midrule
 $\mu T^{\rm full}$                   & $4.1\pm 3.3$  & $4.6\pm 3.5$\\%  &$4.5\pm 3.6$ & $3.8\pm3.6$ & $4.6\pm3.6$ \\
$\mu T^{(1,2)}$  & $5.4\pm 3.6$& $5.9\pm 3.6$\\
$\mu T^{(2,1)}$  & $2.7\pm 3.5$& $3.0\pm 3.5$\\
$\mu T^{\rm combined}$                     & 4.0 $\pm 2.6$& $4.4\pm 2.5$\\
\bottomrule\bottomrule
\end{tabular}
\caption{Values of $\mathcal{D}_\ell^{\mu T} = \ell(\ell+1)\clmut/2\pi$ obtained reanalyzing the \ks $\mu$ distortion maps using their same fiducial mask and power spectrum pipeline. The superscript denotes which data split has been used for each $\mu$ and CMB field respectively. The error bars are based on the jackknife estimate described in Sec.~\ref{sec:khatri-ps} and exceed by $\sim 30\%$ those of \ks. We report the value obtained by cross-correlating different data splits and different component-separated CMB maps. We use the same units of \ks to make facilitate a direct comparison.}
\label{tab:khatri}
\end{table}

\subsection{Foreground contamination}
We test for residual foreground contaminations in the maps using the deprojection techniques introduced in Sec.~\ref{sec:mu-fgdeproj} and adopting the same set of  Galactic and extragalactic foreground templates.  
We first compute the cross-correlation coefficient of both the \planck SMICA DR2 map used in the \ks analysis and their $\mu$ map with all the templates. 
We show that the main foreground residuals in the map are represented by Galactic dust, tSZ and, less importantly but non negligibly, CIB, for which we detect a 60\%, 40\% and 30\% correlation respectively.
The magnitude of the correlation is stable with respect to the choice of different dust (e.g. COMMANDER or GNILC based) or tSZ templates (NILC or MILCA). 
For the SMICA DR2 map, the correlation coefficient we observe with the same set of templates is well below the 5\% on all relevant scales. 
As such, the deprojection applied to the $\clmut$ power spectrum does not bring any statistically significant shift in the first bandpower in particular, and it does not bring the measured power spectra to be consistent with 0.
The sharp oscillatory features suggest that the maps are dominated by primary CMB leakage and the tSZ deprojection seems to enhance those features at small angular scale, pointing to a non-negligible residual in the $\mu$ map. This is somehow expected since the data model of \ks does not include $y$. As such the resulting $\mu$ map is $\sim 30\%$ correlated with the $y$ templates in the first bin, while the SMICA map is anticorrelated with $y$ at the 5\% level on all angular scales. 
When correlating the \ks $\mu$ map with the tSZ-deprojected SMICA map of the \planck DR3 release, which is designed to be tSZ-free, we find consistent results with the analysis discussed above (see Table~\ref{tab:khatri}), supporting the hypothesis of the spectra being dominated by CMB leakage. 

\subsection{Map offsets and CIB monopole uncertainties}
The sky monopole is not constrained by \planck data. 
The correlated noise component in the raw time-ordered-data is in fact modelled by a sequence of baseline offsets of a specific length in time tuned to optimize the noise removal and minimize the computational cost of the map-making step \cite{madam,planck18_lfi,planck18_hfi}. 
Therefore, the monopole cannot be distinguished from a global noise offset and the final map monopoles can assume arbitrary amplitudes.
For \planck LFI instrument, these are estimated during the calibration step and removed from the final map, while for the HFI channels used in \ks, these are fixed to reproduce the value of the dust and CIB monopole. 
While the first set of offset values is estimated from the data through correlation with maps of the HI column density \cite{planck13_hfi}, the latter are inserted according to the expected CIB monopole model of \cite{bethermin2012}. 
The uncertainties on the offsets depend on the specific astrophysical dataset used to constrain them and can be large \cite[see][hereafter O19, and references therein]{odegard2019}.
The recent analysis of the FIRAS team in O19 provided new estimates of the CIB monopole at the \planck HFI frequencies (100, 143, 217, 353, 545, 857 GHz). 
These are the first data-driven values for frequencies below 200 GHz and are sometimes in tension with the model used by \planck for the 217 and 353 GHz channels (we report all relevant values in Table~\ref{tab:cib-monopoles}). 
Despite the component separation algorithm of \ks should remove the nominal monopole injected in the maps, we try to quantify how these uncertainties can affect their final result and its uncertainty. 
For this, we consider a single-pixel toy model where we fit for a $\mu$ distortion in a sky model that includes a CMB temperature anisotropy of $200\,\mu K$ and Galactic dust using only the 100, 143, 217 and 353 GHz as in \ks. 
In the MC simulations, at each frequency we add a random realization of the average \planck noise adopting the numbers in Table 12 from \cite{planck18_hfi} before performing the fit. 
In this case we recover an unbiased estimate of $\mu$. 
We then proceed to perform a similar test where we add to all simulated frequencies not only the instrumental noise but also a systematic error that represents the uncertainty in the map offset after CIB monopole subtraction. 
For each frequency, we draw this error from a Gaussian distribution with standard deviation equal to the error of the measurements of O19 and a mean either equal to 0 or equal to the difference between the CIB monopole values adopted by \planck and those of O19.  These cases cover two different hypothesis: the first accounts only from a data-driven estimate of the statistical uncertainty in the estimate of the CIB offsets residuals, while the second considers also a systematic mean residual in the maps. 

In both cases, the overall error on the recovered $\mu$ (estimated as the standard deviation of the samples) increases compared to the case with no systematic errors. When we introduce a non-zero mean residual offset, the recovered distribution of the $\mu$ pixels becomes inconsistent with zero at $\sim 1.5\sigma$. The recovered CMB also appear slightly biased at a similar statistical significance. 
We repeat the analysis by injecting the systematic errors one frequency at a time and find that the major source of the bias and uncertainty are the monopole errors at 143 and 353 GHz. 
This shows that the limited frequency coverage used in \ks makes their analysis very sensitive to the CIB monopole uncertainties and their statistical error bars underestimate the overall uncertainty in the measurement. 
The ratio between the error bars on $\mu$ obtained including the CIB monopole uncertainties (that we conservatively take as the 68\% upper limit of the recovered $\mu$ distribution) and the error bars derived including only the instrumental noise gives us an estimate of the factor that the measured error bars on $\clmut$ should be inflated by to account for these effects in the final $\fnl$ estimate. This is equal to $\sim 3$ but can get to $\sim 5$ for the case where we consider a systematic mean CIB monopole residual in all frequency bands. We also note that the high sensitivity to the CIB monopoles and related potential biases induced by their misestimation appears only when we fit for the $\mu$ distortion in the frequency range we are considering. In fact, we do not observe any significant bias in the recovered CMB even when systematic errors are injected if we do not include $\mu$ in the data model. Alternative component separation methods working in harmonic domain might be less sensitive to these issues although they might still have an impact through coupling with the analysis mask.
\begin{table}[t]
\begin{tabular}{ccc}
\toprule\toprule
 $\nu$ [GHz]& O19 [MJy/sr]           & Planck [MJy/sr]\\
 \midrule
100&$0.007\pm 0.014$&$0.003\pm 0.003$\\
143&$0.010\pm 0.019$&$0.0079\pm0.0079$\\
217&$0.060\pm 0.023$&$0.033\pm 0.013$\\
353&$0.149\pm 0.017$&$0.13\pm 0.026$\\
%545&$0.371\pm 0.018$&$0.35\pm 0.070$\\
%857&$0.576\pm 0.034$&$0.64\pm0.128 $\\
\bottomrule\bottomrule
\end{tabular}
\caption{Estimates of the CIB monopoles based on \cite{bethermin2012} used to set the zero-level maps of the \planck HFI channels and their new data-driven estimates of \cite{odegard2019}.}
\label{tab:cib-monopoles}
\end{table}

\subsection{Final $\fnl$ constraints}
With the new bandpower estimates provided in Sec.~\ref{sec:khatri-ps} we perform a full likelihood analysis of the single power spectrum bin using the pipeline of Sec.~\ref{sec:fNL_constaints}. 
This includes the full information on the bandpowers and accounts for the scale dependent shape of $\clmut$ and power spectrum binning effects that were neglected in the original \ks analysis. 
Including only the statistical uncertainties, we find comparable constraints to what we obtain with FIRAS at the level of $|\fnl|<3\times 10^{6}$ (see the solid grey line in Fig.~\ref{fig:fNL_constraints}), showing that the original results were underestimated by about an order of magnitude. 
However, if the uncertainties connected with the removal of the \planck HFI maps offsets are accounted for, the constraints significantly degrade and are not competitive with our measurement. 
Specifically, if we only consider the statistical uncertainty in the estimate of the CIB offset residuals and inflate the error bar on $\clmut$ by a factor $\sim 3$, the inferred constraint on $\fnl$ becomes $|\fnl| < 5.5 \times 10^6$.
If we instead also include the systematic mean CIB residual in all frequency bands, and inflate $\Delta\clmut$ by about $5\times$, the $2\sigma$ upper limit loosens to $|\fnl| < 8.6 \times 10^6$ (dashed grey line in Fig.~\ref{fig:fNL_constraints}).

\section{\label{sec:conclusions}Conclusions}

In this paper, we performed the first attempt to self-consistently measure both the monopole and the anisotropic part of the $\mu$-type spectral distortion of the CMB. These provide invaluable information on the thermal history of the universe, on physics beyond the standard model, and on highly-squeezed primordial non-Gaussianities on scales $k \gg 0.05\,\text{Mpc}^{-1}$ that are not constrained by CMB observations both for scalar, tensor, and mixed bispectra of primordial perturbations. 

For this purpose, we used the FIRAS spectrometer legacy data to reconstruct an all-sky map of the $\mu$ distortion. Being FIRAS sensitive to the absolute sky brightness, we managed to reduce the previous constraint on the monopole of $\mu$ distortion by a factor of 2 thanks to a more robust foreground cleaning. We correlated the $\mu$ map with CMB temperature and, for the first time, $E$ and $B$-mode polarization anisotropies measured by the \planck satellite on large angular scales $2\leq \ell\leq 46$. The measured unbinned $\clmut$ and $\clmue$ cross-spectra have been translated to constraints on the amplitude of local-type primordial non-Gaussianity  $\fnl$ at scales $10^2 \lesssim k \lesssim 10^4$ Mpc$^{-1}$ and to an upper limit on its $\nnl$.  Our baseline analysis suggests $|\fnl|<\fnlTE$ from the combined analysis of $\clmut$ and $\clmue$, and $\nnl\lesssim 1.4$ if we assume that $\langle\mu\rangle$ is only sourced by the dissipation of acoustic waves in the primordial plasma. $\clmub$, of which we provide the first constrain, was found to be consistent with zero, implying no strong violation of the statistical isotropy of tensor perturbation. 
We postpone a full analysis and cosmological interpretation of the $\mu B$ cross-correlation to future work.

We have run an extensive suite of systematic checks to assess the robustness of our results against foreground contaminations and analysis choices.   We found that all our results are stable and all the induced shifts are within fluctuations expected from statistical errors. Thanks to their extended spectral coverage and sensitivity to the absolute sky brightness, we show that spectrometers data are very robust against systematic effects. In fact, we re-analyzed previous measurements of $\clmut$ based on data from \planck and found that they are dominated by effects due to an imperfect component separation due to a reduced frequency coverage and by astrophysical uncertainties. Once all these effects are all taken into account those measurements deliver a constraint comparable to ours or worse. Going forward, spectrometers that will be able to measure the $\mu$ distortion monopole accurately are the only way to break the degeneracy with $\fnl$ for the interpretation of $\clmut$ and $\clmue$ in a model independent way. To this end, extending the frequency coverage of spectrometers to $\nu \le 60$ will be crucial as the signatures of $\mu$ distortions are stronger. \\

We plan to make our maps and results publicly available to the community upon acceptance to explore additional theoretical models. \\

As this paper was finalized, we became aware of similar work in \cite{rotti2022nongaussianity} using the full set of \planck channels between $30$ and $857$ GHz. Their $\fnl$ constraint consistently benefits from an improved component separation that reduces foreground contaminations at small scales and allows them to use multipoles $\ell\lesssim 1000$ compared to the  previous \planck analysis by \citet{khatri15}. In addition to providing the first measurements of $\clmub$, our work is complementary to \cite{rotti2022nongaussianity} as it extends the measurements to larger angular scales. 
We note that the $\fnl$ constraints they obtained when considering only angular scales that overlap with this work (which still contribute to $\approx 30\%$ of their sensitivity) are likely optimistic due to a simplistic treatment of the correlation and non-Gaussianity of the noise after component separation, as well as multipole bin-bin correlations in the likelihood. 
A comparison between the theoretical curves adopted in their analysis and the ones used in this work would also be informative.
We expect these effects to push the \planck based constraint from large scale closer to ours. 

\begin{acknowledgments}
We thank Christopher Anderson, Reijo Keskitalo, Daan Meerburg, Niels Odegard, and Giorgio Orlando for useful discussions. We thank Luca Pagano for useful discussion and feedback and his invaluable help with maximum likelihood power spectrum estimation. 
FB would like to thank Simone Aiola and the Center for Computational Astrophysics at Flatiron Institute, where this work was initiated, for hospitality. GF acknowledges the support of the European Research Council under the Marie Sk\l{}odowska Curie actions through the Individual Global Fellowship No.~892401 PiCOGAMBAS. Part of this research used resources of the National Energy Research Scientific Computing Center (NERSC), a U.S. Department of Energy Office of Science User Facility operated under Contract No. DE-AC02-05CH11231.  Some of the results in this paper have been derived using the healpy/\textsc{Healpix} \cite{Zonca2019,healpix} and getdist packages \cite{getdist}, and NumPy \cite{2020NumPy-Array}, SciPy \cite{2020SciPy-NMeth} and Matplotlib libraries \cite{Hunter:2007}.
\end{acknowledgments}

\appendix

\section{Uncertainties on the reconstructed maps}
\label{sec:maps_std}
\begin{figure*}[t]
    \centering
    \includegraphics[width=\textwidth]{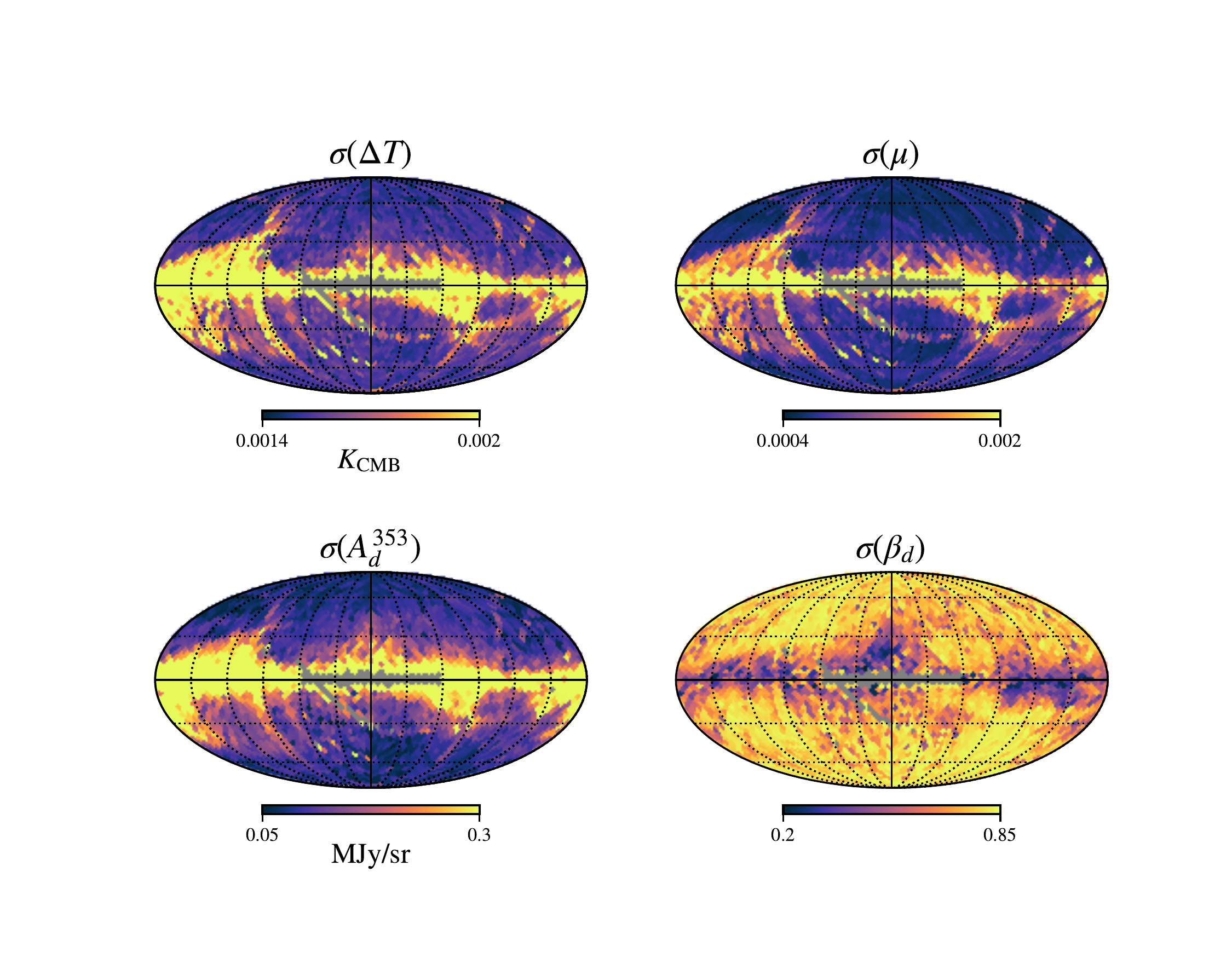}
    \caption{Uncertainties on the reconstructed maps calculated as the standard deviation of the posterior in each pixel. The top left plot shows the CMB dipole, the top right refers to the anisotropic $\mu$-type distortion and the bottom ones show the amplitude and spectral index (at 353  GHz) of the Galactic dust. Pixels greyed out are removed by the FIRAS destriper mask.}
    \label{fig:maps_std}
\end{figure*}

In Fig.~\ref{fig:maps_std} we show the estimated uncertainties on the reconstructed maps in our baseline analysis: $\Delta T$, $\mu$, $A_d$, and $\beta_d$. 
These uncertainties are calculated as the standard deviation of the posterior, obtained through MCMC sampling, in each pixel.

\bibliographystyle{aasjournal}
\bibliography{specdist_firas}% Produces the bibliography via BibTeX.

\end{document}